\documentclass[fleqn,usenatbib]{mnras}

\usepackage{newtxtext,newtxmath}

\usepackage[T1]{fontenc}

\DeclareRobustCommand{\VAN}[3]{#2}
\let\VANthebibliography\thebibliography
\def\thebibliography{\DeclareRobustCommand{\VAN}[3]{##3}\VANthebibliography}

\usepackage{graphicx}	
\usepackage{amsmath}	

\usepackage{comment} 

\title[Stellar bars and galaxy spin]{Measuring the evolution of stellar bars with the host galaxy's spin}

\author[Robin Joshi et al.]{
Robin Joshi,$^{1}$\thanks{E-mail: rjos0964@uni.sydney.edu.au}
Scott M. Croom,$^{1}$
Stefania Barsanti,$^{1}$
Elizabeth J. Iles,$^{1,2}$
Joss Bland-Hawthorn$^{1}$
\newauthor
and Jesse van de Sande$^{2}$
\\
$^{1}$Sydney Institute for Astronomy, School of Physics, A28, The University of Sydney, NSW, 2006, Australia\\
$^{2}$School of Physics, University of New South Wales, NSW, 2052, Australia\\
}

\date{Accepted XXX. Received YYY; in original form ZZZ}

\pubyear{\the\year{}}

\begin{document}
\label{firstpage}
\pagerange{\pageref{firstpage}--\pageref{lastpage}}
\maketitle

\begin{abstract}
We examine to what extent the galaxy spin parameter proxy ($\lambda_R$) is affected by bar formation and how it is related to the strong and weak classifications of stellar bars. By creating mock observations of a simulated galaxy, we show that the emergence of a stellar bar can cause mass-weighted $\lambda_R$ to decrease by up to 16\,\%, depending on the bar's orientation. This decrease can be exaggerated if there is a burst of star formation due to the bar driving gas to the center of the galaxy. We use the SAMI galaxy survey to show that weakly barred galaxies have statistically significant younger average stellar populations, higher galaxy spin proxy and higher specific star formation rates compared to strongly barred galaxies within one effective radius. If we consider galaxies with average light-weighted stellar population age less than 3 Gyr within one effective radius, we still find weakly barred galaxies to have a higher galaxy spin proxy than strongly barred galaxies. Based on these trends found from the SAMI galaxy survey, we suggest weakly barred galaxies are rapidly forming, similar to the bar formation process seen in simulations, while strongly barred galaxies are undergoing slower (secular) evolution.  

\end{abstract}




\section{Introduction}
\label{sec:intro}

Barred galaxies are commonplace in the local Universe. Large surveys of galaxies tell us that somewhere between 25\% to 75\% of spiral galaxies have bars \citep{barfractionmasters2011,barfractionerwin2018}. The bar fraction drops for galaxies at higher redshift \citep{barfracredshift}, but in the era of the James Webb Space Telescope (JWST) we now see barred galaxies at redshifts as high as $z \approx 4$ \citep{JWSTbarfrac2025}. These observations suggest that bars may play an important role in the story of galaxy evolution.

Simulations have been used to study the bar instability for many decades. \citet{oldbarsimhohl1971,oldbarsimostriker1973} were the first to show that cold, rotating disc systems are susceptible to the bar instability. The spontaneous formation of the bar has been commonly attributed to swing amplification and orbit trapping \citep{swinggoldreich1965,toomreswing1966,toomreswing1981,binneyswing2020}. In this paradigm, a small leading spiral, due to the combination of self-gravity, differential rotation and epicyclic motions, will swing into an amplified trailing spiral. This trailing spiral propagates inward and through the center of the galaxy, where it emerges as a leading spiral, restarting the swing amplification process. The bar grows exponentially in strength while this mechanism is efficient.

Isolated galaxy models and simulations also show that bars play a vital role in the secular evolution of a galaxy. For example, bars tend to grow primarily by redistributing angular momentum to resonant orbits in the inner dark matter halo \citep{barangulardeb2000,athabarsecular,bardarkcollier2021,barangkararia2022,baranglu2024joshi}. 
Bars also cause inflows of gas which simulations show can either dynamically heat the gas, thereby preventing star formation \citep{barquenching2018khop} or lead to a starburst at the center of the galaxy \citep{barstarburstcarles2016,lizsimuation2022}.

The growth of the bar is typically quantified by the $A_2/A_0$ parameter, which is a ratio between the $m=2$ to the $m=0$ Fourier amplitude of the surface density measured in azimuth over the radial extent of the bar. Past works using both isolated galaxy and cosmological simulations like \citet{fujii2018fdisk} and \citet{cosmobarfra2021} have used a fixed $A_2/A_0$ value to define whether a galaxy has a bar. However, \citet{joss2023fdisk} showed how fitting an exponential to the $A_2/A_0$ parameter reflects the swing amplification process of bar formation since once the amplitude growth saturates, it leaves the exponential regime and the bar has formed. After the peak amplitude is reached, typically, the bar settles down and undergoes slower (secular) evolution. This raises a question regarding the observations of barred galaxies; at what point does a galaxy appear to be barred and is there a way to distinguish between a forming bar (before peak amplitude) versus a bar in its secular phase (after peak amplitude)?

Galaxies with bars are typically classified into two categories; strongly and weakly barred, ideally at a wavelength unaffected by dust extinction, based on how prominent the bar appears to be. This dates back to the classifications of \citet{weakstrongclassvanc1959} where a spiral galaxy may be barred (SB), non-barred (SA) or somewhere in between (SAB). While this classification scheme is common in the literature, it misses an important characteristic of a stellar bar, i.e., to what extent it has affected the dynamics of the galaxy. 

Attempts have been made to quantify the dynamical effects of the bar with only photometry. For example, \citet{weakstrongbarspincervantes2013} used the model by \citet{galaxyspinmodelmo1998} and the Tully-Fisher relation to estimate galaxy spin defined by \citet{peeblesspin1971} as

\begin{equation}
    \lambda = \frac{L |E|^{1/2}}{GM^{5/2}}
    \label{eq:peeblesspin}
\end{equation}
\noindent where $E$, $M$ and $L$ are total energy, mass and angular momentum for a sample of $\approx$ 10600 galaxies from the Sloan Digital Sky Survey (SDSS). They found that galaxies with short bars tend to have higher galaxy spin than galaxies with long bars. However, this result relies on the accuracy of deriving kinematic quantities like galaxy spin using only photometry. Furthermore, in the absence of kinematic data, no concrete conclusions can be made regarding how the evolution of a bar changes galaxies dynamically.

Despite the shortcomings of morphological bar classifications, previous studies have shown some differences between the two bar populations. Generally, the presence of a bar in a galaxy has been linked to reduced star formation activity \citep{barquenchgalaxy2018,barquenchgalaxy2020}. Strongly barred galaxies in the SDSS catalog have been found to have lower star formation activity and host older, redder populations of stars than weakly barred and unbarred galaxies \citep{weakstrongbarstarformvera2016}. \citet{weakstrongbarcontgeron2021} further analyzed a sample of $\approx$ 1800 galaxies from Galaxy Zoo DECaLS to find that only strongly barred galaxies quench star formation through secular evolution. Despite this difference in the two populations of galaxies, once they account for bar length, any difference in star formation between weak and strong bar galaxies disappears, suggesting that weak and strong bars are the same phenomena on a continuum.

A robust alternative to the visual classification of galaxies is to use the kinematic properties of stars within a galaxy. In the age of integral-field spectroscopy, the Sydney-AAO Multi-object Integral field spectrograph (SAMI) \citep{sami2012} and the Mapping Nearby Galaxies at Apache Point Observatory (MaNGA) instrument \citep{Manga2015} have produced a large sample of galaxies with stellar kinematic measurements. Recently, kinematic measurements from MaNGA have been used on a sample of 225 barred galaxies by \citet{strongweakkinematicgeron2023}, who find weak bars to have slightly higher pattern speeds than strong bars, albeit with significantly overlapping distributions. However, the pattern speeds were measured with the Tremaine-Weinberg method \citep{tremaineweinberg1984} which only gives accurate measurements at certain inclination and bar angle ranges \citep{twmethodzou2019}.

One way to track the effect of the bar on its host galaxy is to track how the galaxy has changed dynamically in the inner regions of the galactic disc. This can be quantified by using stellar kinematic measurements such as the galaxy spin proxy
\begin{equation}
    \lambda_R = \frac{\langle R|V| \rangle}{\langle R \sqrt{V^2 + \sigma^2} \rangle}
    \label{eq:galspinnoweight}
\end{equation}
\noindent where $V$, $R$ and $\sigma$ are the velocity, radius and velocity dispersion. We refer to the galaxy spin proxy as the galaxy spin from this point onward. This was first introduced by \citet{2007spinparameter} as a way to quantify the angular momentum within a given radius (typically the effective radius) or the fraction of dynamical support by rotation versus random motion. \citet{harbornespin2019} showed using a suite of N-body galaxy models that the theoretical relationship of $\lambda \approx \sqrt{2}/3 \lambda_R$, where $\lambda$ can be the spin parameter defined by \citet{peeblesspin1969} (see Eq.~\ref{eq:peeblesspin}) or the alternative by \citet{bullockspin2001}, holds as long as corrections due to observational biases are taken into account. This parameter was initially used to classify early-type galaxies into two distinct categories: slow and fast rotators \citep{edgeonspineric2011}. Eventually late-type galaxies were also included in these samples to better understand how the dynamics of galaxies changes with mass, environment and other parameters \citep{jessespin2017,jesseaperature2017,spinlatetypesande2021}. However, \citet{croomAgespin2024} found that galaxy spin primarily correlates with stellar population age. In fact, after accounting for the stellar population age across the entire SAMI sample, there is little to no trend with mass or environment. A possible contributor to the spin-age relation is the internal evolution of a galaxy; particularly how the bar dynamically heats discs over time, both during its formation and its subsequent evolution. Since simulations of disc galaxies show how bars play a prominent role in the redistribution of angular momentum, one would expect galaxy spin to change between populations of strong and weak bars.

In this study our goal is to explore how bar formation and subsequent evolution effects galaxy spin and how galaxy spin is related to the weak and strong classifications of barred galaxies. We make use of simulations, mock observations and a sample of galaxies from the SAMI galaxy survey to explore this question. In Sec.~\ref{sec:data} we present details on our simulation, data and methods used. Sec.~\ref{sec:results} we present our results and in Sec.~\ref{sec:discussion} provide a discussion. Sec.~\ref{sec:conclusion} presents a summary of our findings.

\section{Data and Methods}
\label{sec:data}

\subsection{Simulations}
\label{sec:simulations}

Detailed information on the simulation we use in this study can be found in \citet{lizsimuation2022,ilessimualtion2}, however for completeness we provide a brief description. A simulation of an isolated stellar disc with a live dark matter halo, stellar bulge and gas disc is used to examine the change in the kinematic properties of a disc galaxy due to the formation of a bar. Known as IsoB in \citet{lizsimuation2022,ilessimualtion2}, this simulation is set up to match the observational properties of NGC 4303, which is reflected by the isolated evolutionary history and final morphology. 

The initial conditions were produced using the \textsc{galic} package \citep{galicinitial2014}, motivated by the kinematic and surface density profile of NGC 4303. The exponential stellar disc is of mass $2.611 \times 10^{10}M_\odot$ and scale length $2.060$ kpc. The stellar bulge follows a Hernquist profile and is of mass $4.02\times10^9M_\odot$ with scale length $2.057$ kpc and the NFW-like dark matter halo is of mass $3.716 \times 10^{11}M_\odot$ and scale length $20.57$ kpc. Finally, the exponential gas disc has a mass resolution of $1044M_\odot$ with total mass $5.22\times10^{9}M_\odot$ and scale length 3.090 kpc. 

The realization of this galaxy configuration starts with a total of $1\times10^{6}$ star particles in the stellar disc, $1\times10^5$ star particles in the bulge and $5\times10^6$ gas particles. The evolution of this galaxy for 1 Gyr was performed using the \textsc{gasoline2} smoothed particle hydrodynamics (SPH) code of \citet{gasoline2wadsley2017}, with their prescribed standard hydrodynamical treatment of 200 neighbours and a Wendland C4 kernel \citep{wendlandkernel2012}. The softening length for stars, dark matter halo and gas are 0.05 kpc, 0.1 kpc and 0.01 kpc, respectively. Star formation occurs if there is convergent flow and the temperature threshold of 300 K and density threshold of 100 atom cc$^{-1}$ conditions are met. The star formation efficiency is set to 10\% with the \citet{imf2003} initial mass function. Photoelectric and UV heating along with metal cooling is implemented as a tabulated cooling function \citep{coolingshen2010}. Stellar feedback from supernova is modeled as superbubbles described in \citet{superbubble2014}. It is important to note that the gas component in this model is singular, in the sense that different gas components, such as atomic and molecular are not differentiated.  

Recent work by \citet{fujii2018fdisk} and \citet{joss2023fdisk} suggests that the primary predictor of the timescale of bar formation in an isolated simulation is the degree of dominance of baryons over the dark matter halo in the inner regions of the galaxy. This is quantified by the parameter $f_{\text{disc}}$, defined as 

\begin{equation}
    f_{\text{disc}} = \left( \frac{V_{c,\text{disc}}(R_s)}{V_{c,\text{tot}}(R_s)} \right)_{R_s = 2.2R_d} 
    \label{eq:fdisk}
\end{equation}

where $V_c(R)$ is the circular velocity at radius $R$. There are also second order effects that can change the bar formation timescale, namely the Toomre Q parameter and scale height \citep{chenshen1} along with the spin of the dark matter halo \citep{chenshen2}. For now we consider only the primary predictor, $f_{\text{disc}}$. In this simulation $f_{\text{disc}} \approx 0.66$ with a gas mass fraction $M_{\text{gas}}/M_{\text{disc}} = $  0.2. As we explore in Sec.~\ref{sec:barformationproperties}, the bar forms in around 400 Myr which is about the timescale predicted by the results of \citet{joss2023fdisk}.

\subsection{Generating Observables with \textsc{simspin}}
\label{sec:simspin}

A key part of our exploration revolves around how different projections of a barred galaxy may effect $\lambda_R$ along with the effect of weighting $\lambda_R$ by flux. To do this we make use of \textsc{simspin}

\citep{simspin2019paper1,simspin2020paper2,simspin2023paper3}, which can take an isolated simulation and generate flux maps along with providing the mass-weighted line of sight (LOS) velocity and LOS velocity dispersion. The alignment of the galaxy is determined through the process described in \citet{alignment2019}, where the eigenvalues and eigenvectors of the reduced inertia tensor of the half mass ellipsoid of stellar particles are calculated. Using \textsc{prospect} \citep{prospect2020}, the simple stellar population model (SSP) \textsc{galaxev} \citep{ssp2003} is used to create a spectral energy distribution for each star particle based on the age, metallicity and birth mass of each star particle from our simulation. The initial condition star particles are given an age of 3 Gyr at the start of the simulation, which get older and reach an age of 4 Gyr at the last snapshot of the simulation. While the resulting age distribution is idealized compared to a real galaxy, this configuration allows us to clearly analyze the effect newly formed stars may have on galaxy spin measurements, although we acknowledge any effect we see may be exaggerated due to this choice. 

\textsc{simspin} gives users the option of choosing the telescope with which to make the mock observations, along with the orientation and distance of the galaxy. Our goal with creating these mock observations is to explore how the orientation of the bar can effect measurements of stellar kinematics. Therefore, to make sure our results are not affected by spatial resolution, our mock observations are made with the MUSE telescope setting which provides a spatial sampling of 0.2 arcsec/pixel. The galaxy at each snapshot is put at a distance where 1\,arcsec corresponds to 1 kpc. A Gaussian point spread function with full width half max 0.8 arcsec is used to convolve the spatial plane to include seeing which only affects the flux. The observed wave range provide by this setting is 4700.15–9351.4 \r{A}. Information related to the chosen inclination and alignment of the galactic bar is given in Sec.~\ref{sec:galaxyspinbarform}.

\subsection{SAMI Data}
\label{sec:samiexplain}

Information on the SAMI instrument and the galaxy survey can be found in \citet{sami2012, samitarget2015}. In brief, SAMI was equipped with 13 hexabundles \citep{hexabundle2011} which were deployed over a 1$^\circ$ field of view. Each hexabundle covered $\approx 15$ arcsec diameter region with 61 individual 1.6 arcsec fibres. All 793 object fibres and 26 sky fibres connect to the AAOmega spectograph \citep{aaospectograph2006}, which has a blue (3750 - 5750 \r{A}) and red arm (6300 - 7400 \r{A}). The blue arm 580V grating gives spectral resolution $\approx 1810$ at 4800 \r{A} while the red arm 1000R grating gives spectral resolution $\approx 4260$ at 6850 \r{A} \citep{spectralresolution2018}.  

The main measurement we explore in our study is the galaxy spin within one effective radius ($\lambda_{Re}$). This is measured using SAMI spatial kinematic maps in \citet{jessespin2017}, with an aperture correction applied by \citet{jesseaperature2017} and a further seeing correction from \citep{kateseeing2020,jesseseeingspin2021}. To account for the affect of inclination on $\lambda_{Re}$, an edge-on spin $\lambda_{Re_{EO}}$, which is the expected galaxy spin if the galaxy was orientated edge-on, is calculated. In brief, the starting point is the relationship between observed ellipticity ($\epsilon_e$) and intrinsic ellipticity ($\epsilon_{\text{intr}}$) given by \citet{kinematicsreview2016}

\begin{equation}
    \epsilon_e = 1 - \sqrt{1+\epsilon_{\text{intr}}(\epsilon_{\text{intr}}-2)/\sin^2i}
\end{equation}

\noindent where $i$ is the inclination of the galaxy. For different inclinations, the observed $(V/\sigma)_{\text{obs}}$ and edge-on $V/\sigma$ are related by

\begin{equation}
    \left(\frac{V}{\sigma}\right)_{\text{obs}} = \left(\frac{V}{\sigma}\right) \frac{\sin i}{\sqrt{1-\delta\cos^2i}}
\end{equation}

\noindent $\delta$ related to the anisotropy parameter \citep{cappellariedgeon2007,edgeonspineric2011,edgeonspinkate2019}. The value of $\delta$ will depend on $\epsilon_{\text{intr}}$ of the galaxy. Finally, the edge-on $V/\sigma$ can be used to find $\lambda_{Re_{EO}}$ as outlined in \citet{2007spinparameter,edgeonspineric2011}

\begin{equation}
    \lambda_R = \frac{\kappa (V/\sigma)}{\sqrt{1 + \kappa^2(V/\sigma)^2}}
\end{equation}

\noindent where $\kappa$ is estimated to be 0.97 from \citet{jesseaperature2017}. Note that the correction is not applied to galaxies that fall outside the model (see \citet{spinlatetypesande2021} for more details). From this point onward we refer to $\lambda_{Re_{EO}}$ as $\lambda_{Re}$ when referring to galaxy spin measurements from SAMI. 

The other parameters we examine are stellar mass, mean stellar population age and the star formation rate within one effective radius ($R_e$). SAMI data release 3 \citep{samidr32021} provides both stellar mass of each galaxy and the star formation rate. Stellar mass is drawn from the Galaxy and Mass Assembly (GAMA) survey \citep{gama2011} and from \citet{samiclustermass2017} for galaxies in cluster regions. The star formation rates within 1 $R_e$ are derived from the extinction corrected H-$\alpha$ flux and the \citet{samisfrrelation1994} relation with a \citet{imf2003} initial mass function correction. Finally, the mass and light-weighted stellar population ages comes from \citet{stellarpops2022}, who use the penalized-pixel fitting code \textsc{ppxf} \citep{ppxf2004,ppxf2017} to fit the MILES SSP \citep{MILES2017} to the spectrum from Voronoi binned \citet{voroni2003} flux maps.   

The properties described above are compared in a statistical sense in Sec.~\ref{sec:samidataresults} between three populations; strongly barred, weakly barred and non-barred galaxies. The classifications are from A. Fraser-McKelvie (private communication), who combined visual classifications from four experts, details of which can be found in \citet{barcatalogue2024}. In this work, a bar is classified as strong if there is a vote fraction of at least 1/3 in this category. A bar is classified as weak if there is a vote fraction of at least 1/3 and it has not been classified as strong already. Any galaxy with a vote fraction of at least 1/3 in the boxy-peanut (BP) bulge category has been classified as strongly barred, since this structure typically emerges after bar formation \citep{bpbulgesim2006,bpforamtion22020} and the orbits supporting these structures are three dimensional bar orbits \citep{bpbulgeorbits2015}. Starting with 3071 galaxies, we only include those galaxies in our sample that have measurements for all of the parameters discussed above (removing 479 galaxies). Furthermore, we remove 177 slow rotators along with 817 galaxies with stellar mass below $10^{9.5} M_\odot$ due to the incomplete kinematic sample below this stellar mass \citep{jesseseeingspin2021}. To further ensure the purity of our sample we use the SAMI data release 3 visual catalogue \citep{lucamorpho2016,samidr32021} to remove 408 galaxies classified as being in either the elliptical or elliptical/S0 categories. This brings our sample to 338 strongly barred, 225 weakly barred and 627 non-barred galaxies. Of the 363 strongly barred galaxies, 104 were classified as having BP bulges.

\section{Results}
\label{sec:results}

\subsection{Bar formation and properties}
\label{sec:barformationproperties}

\begin{figure*}
	\includegraphics[width=\textwidth]{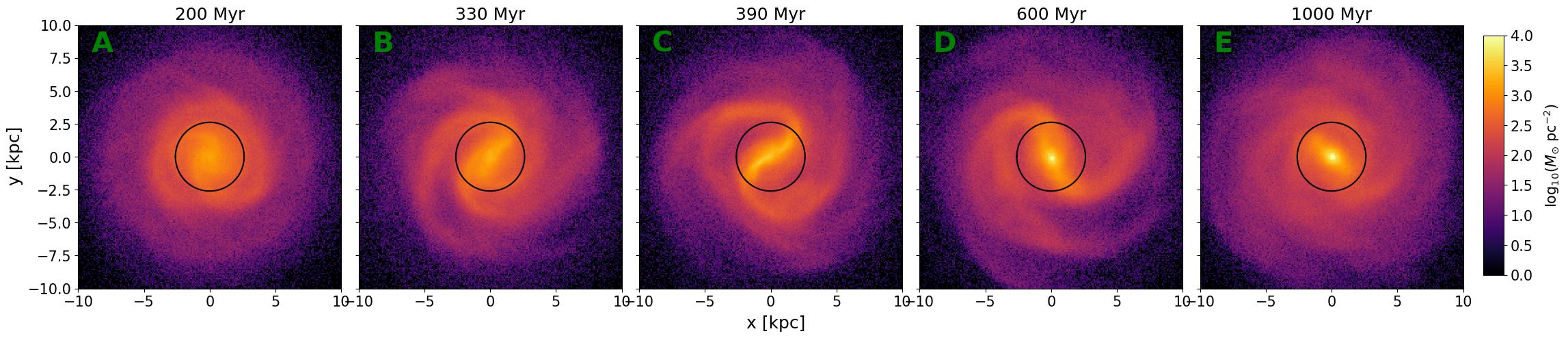}
    \caption{Face-on surface density maps of our simulation depicting different stages in the evolution of the bar. Panel A: the disc before bar formation. Panel B: when $A_2/A_0$ first exceeds 0.2. Panel C: when the exponential growth of $A_2/A_0$ ends. Panel D-E: evolution of the disc after bar formation. The black circle shows where we define the bar radius.}
    \label{fig:faceon_maps_sim}
\end{figure*}

\begin{figure}
	\includegraphics[width=\columnwidth]{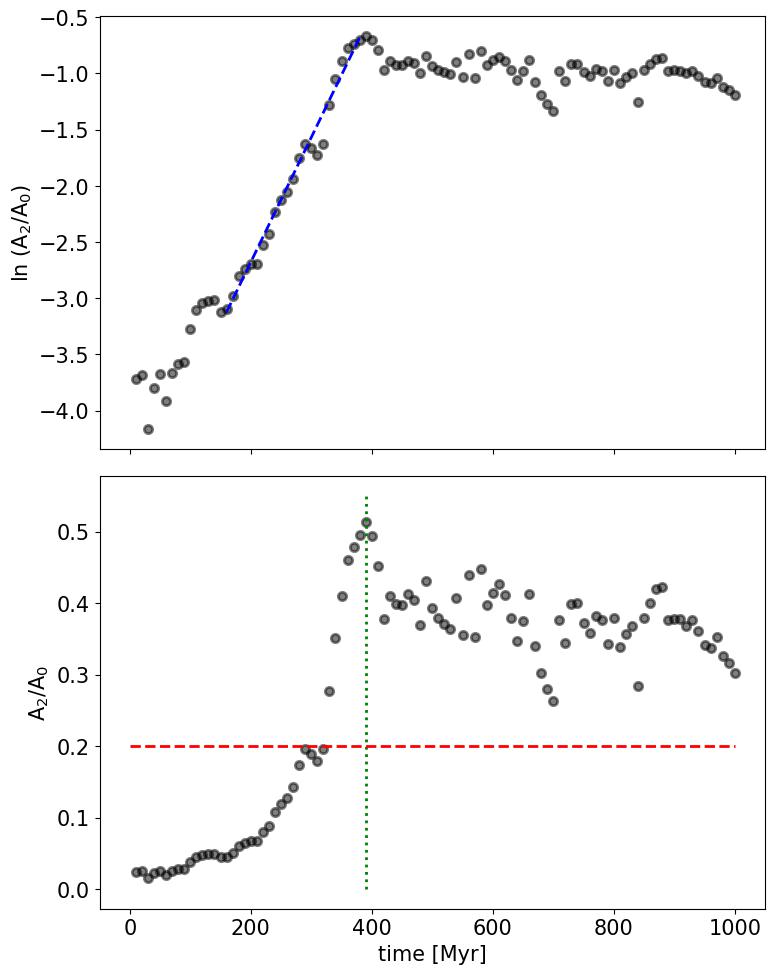}
    \caption{Top panel: Bar strength on a natural log scale. The dotted blue line shows the fit while the bar forms undergoing exponential growth. Bottom panel: Bar strength on a linear scale. The dotted green line shows when the bar strength peaks, or when the bar matures, while the dashed red line indicates $A_2/A_0 = 0.2$.}
    \label{fig:bar_strength_sim}
\end{figure}

We first examine the formation of the bar along with the characteristics of the bar over the course of the simulation. Fig.~\ref{fig:faceon_maps_sim} shows the face-on log scaled surface density at different times during the sequence of bar evolution. Visually, we see that the bar persists over the course of the simulation after its formation while the prominent spiral arms seen at 390 Myr lose their strength in amplitude by the end of the simulation at 1000 Myr.  

To examine in more detail the bar formation period seen in the second and third panels of Fig.~\ref{fig:faceon_maps_sim}, we track the evolution of bar strength in Fig.~\ref{fig:bar_strength_sim}. To do this we calculate the Fourier coefficients 

\begin{equation}
    \frac{A_2}{A_0}\left(r \right) = \left| \frac{ \sum_{j \in \alpha} \mu_je^{i2\phi_j}}{\sum_{j \in \alpha}\mu_j}\right|
    \label{eq:barstrength}
\end{equation}

\noindent where particle mass ($\mu$) and azimuthal angle ($\phi$) are binned in $\approx 0.1$ kpc radial bins ($\alpha$). We take the maximum of Eq.~\ref{eq:barstrength} at each timestep to measure the bar strength, as shown in the bottom panel of Fig.~\ref{fig:bar_strength_sim}.

Following \citet{joss2023fdisk}, we recognize the exponential growth of this parameter as the period when bar formation occurs through the swing amplification mechanism. We find that the bar fully forms at 390 Myr (a face-on surface density map of this time is shown in the third panel of Fig.~\ref{fig:faceon_maps_sim}), with an exponential growth rate of about 11.1 Gyr$^{-1}$. This growth rate is found by fitting a line to the natural logarithm of the bar strength, shown by the dotted blue line in the top panel of Fig.~\ref{fig:bar_strength_sim}. There is an settling time of $\approx$ 100-150 Myr at the beginning of the simulation, which explains why the earlier time-steps do not tightly lie on the fitted line (also mentioned in \citet{lizsimuation2022}). Therefore, we see the swing amplification mechanism begins at about 160 Myr and remains efficient until 390 Myr. A vertical, dotted green line in the bottom panel of Fig.~\ref{fig:bar_strength_sim} is drawn at 390 Myr, which is when we define the bar to be formed, after which the bar undergoes secular evolution.

As mentioned in Sec.~\ref{sec:intro}, it is common to consider a galaxy barred if $A_2/A_0$ exceeds 0.2. This is shown by the red, dashed line in the bottom panel of Fig.~\ref{fig:bar_strength_sim}. The bar reaches this threshold at 330 Myr in our simulation. A face-on surface density map of this time is shown in the second panel of Fig.~\ref{fig:faceon_maps_sim}. Compared to when the bar saturates at 390 Myr, the bar at 330 Myr is quite weak, but still discernible. Hence, one can ask whether bars observed to be weak are rapidly growing due to the formation process, while strong bars are undergoing secular evolution. This is something we explore in Sec.~\ref{sec:strongweakexplain}. 

To understand the difference between the bar before and after saturating (peak $A_2/A_0$), we must know the bar's properties after forming. After forming, the bar has a pattern speed ($\Omega_p$) that does not change significantly. To find $\Omega_p$ we use the particles within 2$R_d$ to calculate a single $m=2$ Fourier mode at a given time-step, find the phase angle and then find the change in this angle over time. We find this results in a scatter of values which does not show any major decreasing trend. Hence, we report the pattern speed $\Omega_p \approx 52 \pm 4$ km s$^{-1}$ kpc$^{-1}$. This puts the average corotation radius of the bar at around 3.37 $\pm$ 0.09 kpc, which was found by determining the radius where $\Omega_p$ equals the angular frequency of the disk at each timestep. 

Next, we measure the length of the bar. One of the ways \citet{athabarvolume2002} measure the length of a bar is to find where difference between the projected face-on surface density along the major and minor axes of the bar reaches zero. Hence, if we take the surface density of the bar along the major and minor axes, then the non-zero radius where the difference between these surface densities is zero is where the bar ends. Practically, for a simulation, we take the radius where the difference is two percent of the maximum. The tolerance value of two percent is arbitrary, and is chosen as it provides a reasonable estimate for the measurement. Higher tolerance values would decrease the measured bar length, while lower tolerance values might increase the bar length unrealistically due to noise in the outer disk.

Similar to $\Omega_p$, we find that there is a scatter of values with no obvious trend, so the bar in our simulation has radius $2.61 \pm 0.61$ kpc after forming. It is worth noting that due to the spiral arms in this simulation, the bar length may be an overestimate, however we find similar values of bar length using Fourier based methods, so we do not anticipate this to change our findings.

The bar slowdown and increase in bar length is associated with the movement of angular momentum from the inner to the outer disc and dark matter halo \citep{athabarsecular}. As our simulation does not run for a long time after bar formation, it is reasonable for the bar to not change significantly after forming, hence the unchanging pattern speed and length. 

\subsection{Dynamical heating and star formation due to bar formation}
\label{sec:heatingsfr}

\begin{figure}
	\includegraphics[width=\columnwidth]{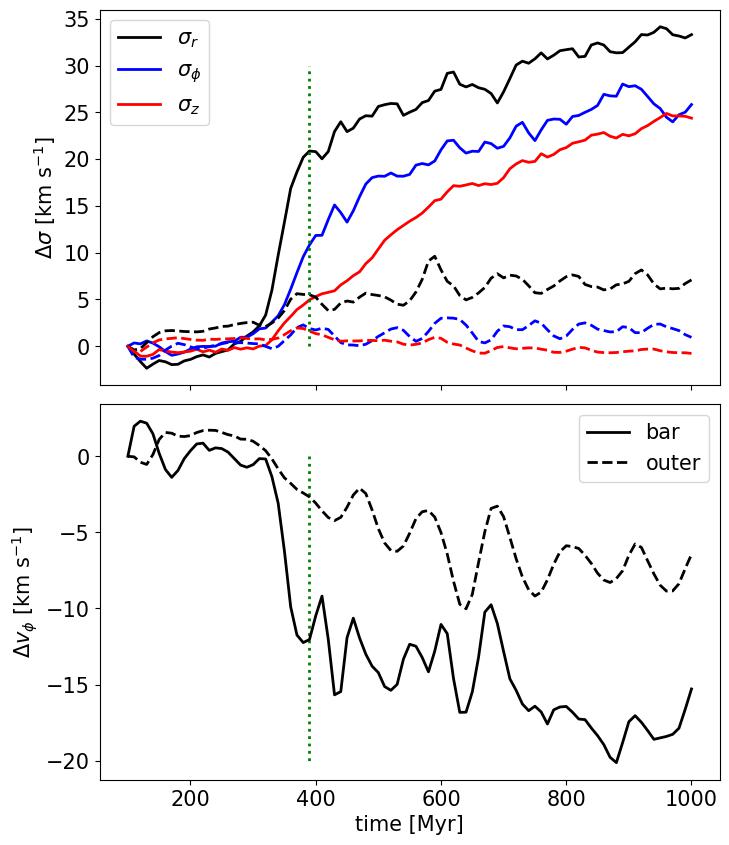}
    \caption{Evolution of the kinematic properties in our simulation. The dotted, vertical green line shows the formation time of the bar. Solid lines show the evolution of the indicated quantity within the bar radius and dashed lines outside the bar radius. Top panel - the evolution of the change in velocity dispersions ($ \sigma_i(t) - \sigma_i(100)$) in each region. Bottom panel - evolution of change in mean azimuthal velocity in each region ($ v_\phi(t) - v_\phi(100)$).}
    \label{fig:vel_disp_sim}
\end{figure}

\begin{figure}
	\includegraphics[width=\columnwidth]{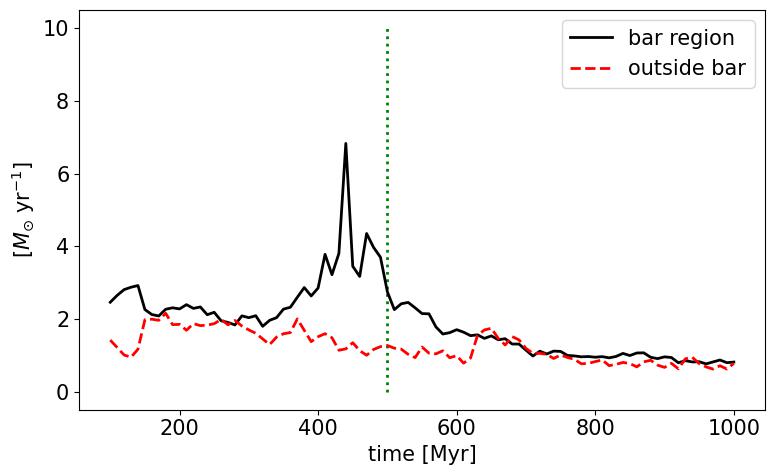}
    \caption{Star formation rate inside the bar region is shown by the solid black line while outside the bar region is shown by the dashed red line. The dotted, vertical green line shows the bar formation time.}
    \label{fig:agesfr_sim}
\end{figure}

It is hard to determine the kinematic history of a galaxy if only a snapshot is given, as is the case with extragalactic observations. However, statistical trends found using large galaxy surveys can be interpreted accurately as long as the physics that underlies galactic evolution are understood. Since our goal is to understand the role of bar formation in the evolution of a galaxy, in this section we look at some possible kinematic signatures that are associated with bar formation. To do this we examine the evolution of the velocity dispersions and azimuthal velocity. Our analysis is done in two regions of the stellar disk, inside and outside the radius defined by the length of the bar (Sec.~\ref{sec:barformationproperties}). 

The top panel of Fig.~\ref{fig:vel_disp_sim} shows the change in velocity dispersion since the 100 Myr snapshot. We choose 100 Myr as the reference since the disc undergoes a settling period at times earlier than this. The sharp increase in $\sigma_r$ within the bar radius is immediately obvious, which increases by about 20 km s$^{-1}$ at the time of bar formation. We also see an increase in $\sigma_\phi$ and $\sigma_z$ inside the bar radius, however these increases occur at slightly different times. The increase of dispersion inside the bar volume in each direction corresponds to a process associated with the formation of the bar. Radial dispersion increases as more particles are caught in the potential of the bar, making their orbits more elliptical. Azimuthal dispersion increases after the rapidly forming bar begins to redistribute angular momentum from the bar volume to other parts of the galaxy. The movement of angular momentum is also reflected in the decrease of mean azimuthal velocity in the same region by about 13 km s$^{-1}$ at bar formation, which is seen in in the bottom panel of Fig.~\ref{fig:vel_disp_sim}. Finally, the vertical dispersion increases due to the gradual formation of the boxy-peanut bulge. 

Fig.~\ref{fig:vel_disp_sim} also shows the change in dispersion and mean azimuthal velocity in the outer disk (dashed lines). While the azimuthal and vertical dispersions remain relatively constant over the course of the simulation, the radial dispersion does increase by about 5 km s$^{-1}$. The mean azimuthal velocity does also decrease by less than 10 km s$^{-1}$ by the end of the simulation. Both of these changes are likely due to the spiral arms which are prominent for a few hundred million years after bar formation. 

Next, we briefly examine how star formation is affected by bar formation. The bar funnels gas to the center of the galaxy, which causes a burst of star formation \citep{barstarburstcarles2016}. Fig.~\ref{fig:agesfr_sim} shows the evolution of star formation rate calculated with the stars formed in the last 10 Myr at each timestep. The key event occurs shortly after bar formation, where there is a sharp increase in star formation inside the bar radius. After the starburst, the star formation rate in the bar region falls to less than half the rate before the formation of the bar. The outer region does not show any significant star forming events, indicating that the increase in star formation is driven primarily by the bar.

\subsection{Bar torque - a case for the spin parameter}
\label{sec:torque}

\begin{figure*}
	\includegraphics[width=\textwidth]{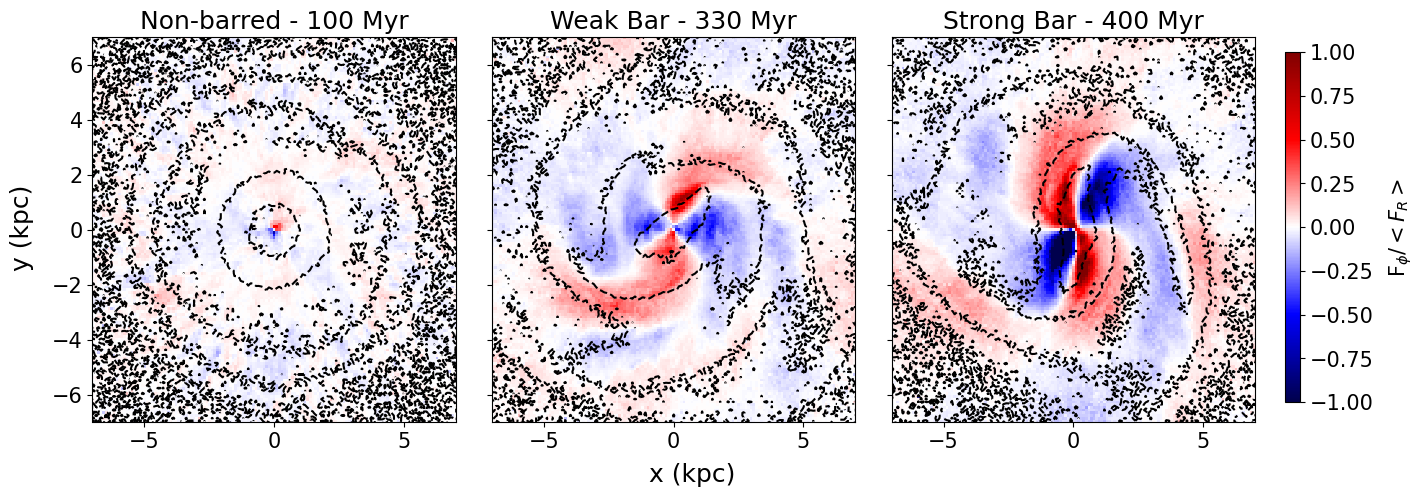}
    \caption{The ratio of tangential to radial force before bar formation (left), during bar formation (middle) and after bar formation (right). The black contours outline the mass distribution at each timestep.}
    \label{fig:bartorque}
\end{figure*}

In Sec.~\ref{sec:barformationproperties} we identify snapshots where the bar is weak and where the bar is strong. However, \citet{butabartorque2001} and \citet{davistorque2025} discuss how bar torque is preferable to $A_2/A_0$ as a measure of bar strength; since the $A_2/A_0$ parameter depends on surface density profiles it only indirectly sees how a galaxy is being affected dynamically. 

To determine how a strong and weak bar identified in Sec.~\ref{sec:barformationproperties} affects the disc, we determine the ratio of tangential force ($F_\phi$) to the magnitude of the radially averaged radial force ($\langle F_R \rangle$) in Fig.~\ref{fig:bartorque}, similar to the $Q_T$ parameter in \citet{combesbartorque1981} and \citet{butabartorque2001}. Before the bar forms the average magnitude of the ratio inside the bar radius is about 0.02. While the bar is forming, or when the galaxy is weakly barred, we see a quadrupole pattern with average magnitude 0.19 in the bar region. After the bar has formed, or when the bar is classified as strong, the quadrupole pattern increases in magnitude with the average around 0.39 within the bar region. In Fig.~\ref{fig:agesfr_sim}, the large burst of star formation occurs only after the bar has entered its secular evolution stage, suggesting that the torque of a bar undergoing rapid formation is not strong enough to dramatically change characteristics such as star formation rate rapidly.  

Galaxy spin quantifies the kinematics within the inner regions of a galaxy. Therefore, if the bar is the primary driver of evolution within the measured region, then it gives an idea of how much the formation and subsequent evolution of the bar has affected the galaxy. One can predict a weak, rapidly forming bar to have a higher galaxy spin than a formed strong bar.

\subsection{Changes to the galaxy spin due to bar formation}
\label{sec:galaxyspinbarform}

Our goal is to understand the how bar formation affects the observed kinematics of galaxies. The bar plays a major role in redistributing the angular momentum from the inner disc to the dark matter halo, as discussed in Sec.~\ref{sec:intro}. If the galaxy spin of \citet{2007spinparameter} reflects the kinematics within a given region of the disc, then the formation of the bar should cause a change in this parameter. In practice, Eq.~\ref{eq:galspinnoweight} is defined by

\begin{equation}
    \lambda_R = \frac{ \sum_{i=1}^NX_iR_i|V_i| }{\sum_{i=1}^N X_iR_i \sqrt{V_i^2 + \sigma_i^2} }
    \label{eq:galspinweight}
\end{equation}

\noindent where the sum is over bins and $X$ is the weight assigned to each bin. The radius $R_i$ is the semimajor axis of the ellipse on which bin $j$ lies. Typically $X$ is the flux from a given bin, but to make a comparison between our simulation and the r-band flux given by \textsc{simspin} (details given in Sec.~\ref{sec:simspin}) we also calculate galaxy spin using the stellar mass in each bin. 

\begin{figure}
	\includegraphics[width=\columnwidth]{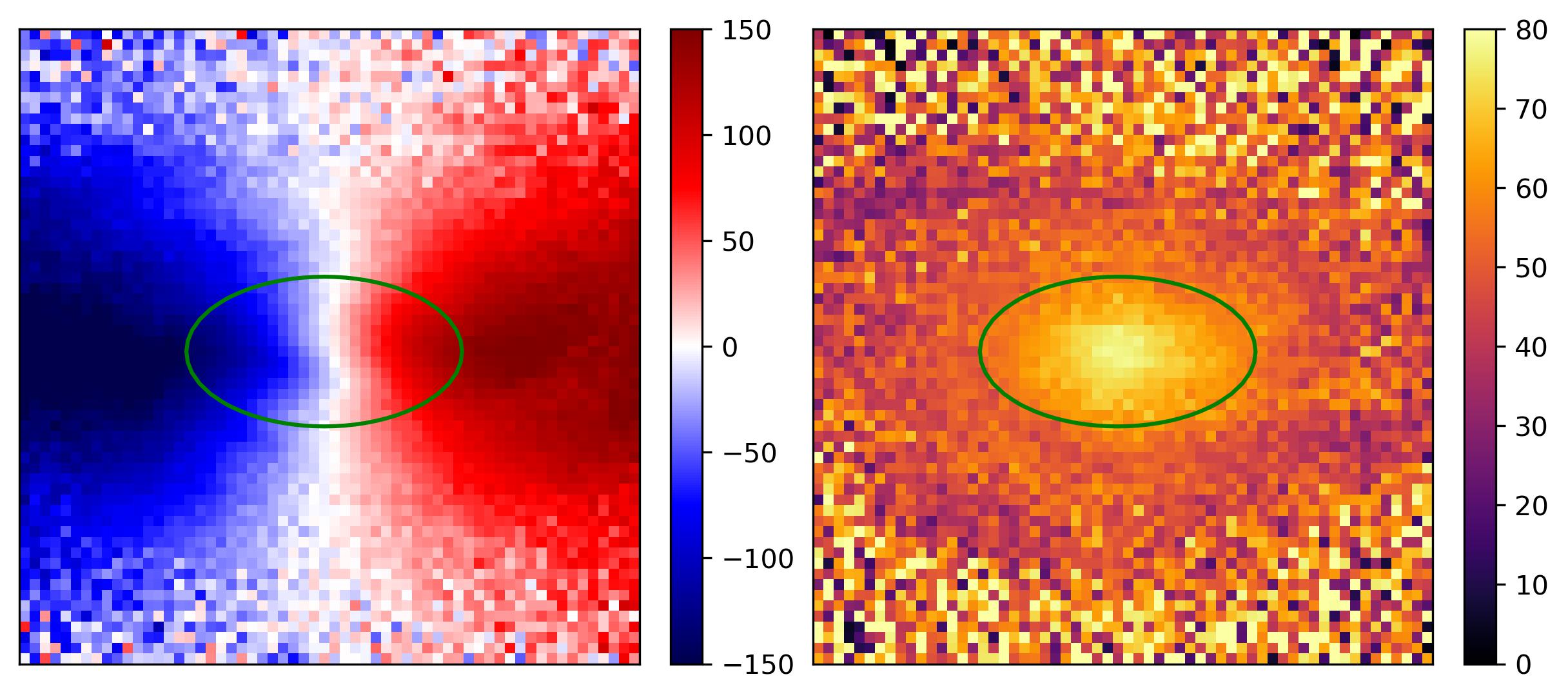}
    \caption{Examples of the line of sight velocity (left panel) and the line of sight velocity dispersion (right panel) of the 330 Myr snapshot with $i = 60^\circ$ and the weak bar aligned with the horizontal axis. The green ellipse indicates the measurement ellipse used for this galaxy orientation in our study.}
    \label{fig:velmapexample}
\end{figure}

\begin{figure}
	\includegraphics[width=\columnwidth]{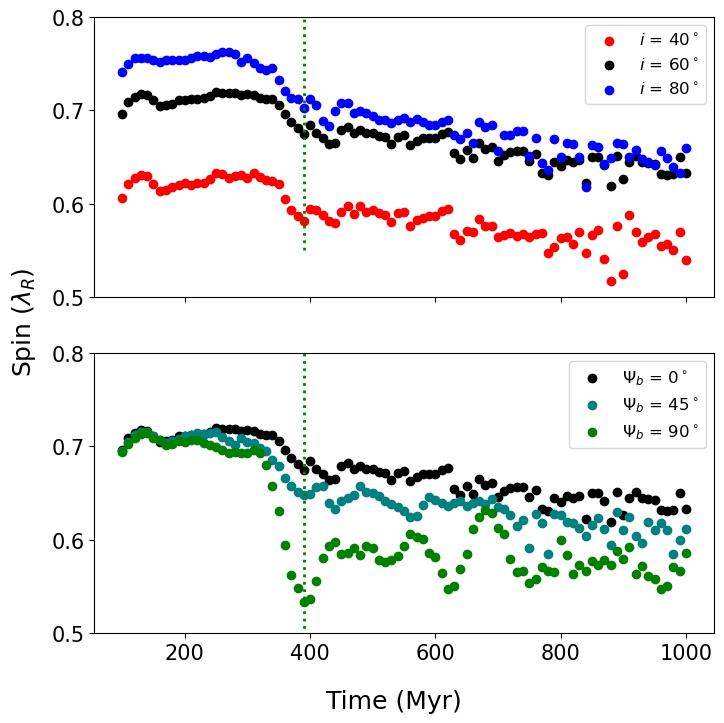}
    \caption{The evolution of mass weighted galaxy spin within a 1 $R_e$ ellipse measured at t = 200 Myr. Top panel - at inclinations of $i = 40^\circ$, $i = 60^\circ$ and $i = 80^\circ$ with the bar aligned with the horizontal axis. Bottom panel - at a fixed inclination of $i = 60^\circ$ with the orientation of the bar with respect to the horizontal axis at $\psi_b = 0^\circ$, $\psi_b = 45^\circ$ and $\psi_b = 90^\circ$. The dotted green line indicates when the bar forms. }
    \label{fig:massspin}
\end{figure}

We create mock observations of our simulation at every timestep with \textsc{simspin} to determine if $\lambda_R$ changes due to bar formation and if this change is dependent on the orientation of the galaxy or bar. An example of the line of sight velocity and velocity dispersion maps are shown in Fig.~\ref{fig:velmapexample}. In the top panel of Fig.~\ref{fig:massspin} and Fig.~\ref{fig:fluxspin} (discussed in Sec.~\ref{sec:fluxspinbarform}), the evolution of $\lambda_R$ is calculated at inclinations ($i$) of 40$^\circ$, 60$^\circ$ and 80$^\circ$ with the bar aligned with the horizontal axis. We calculate galaxy spin at each timestep within the same half mass radius ellipse, found with the 200 Myr timestep using a multi-Gaussian expansion (MGE) \citep{mgefit2002} of the mass distribution. While this approach is simplistic, as one would expect the evolution of the galaxy to change the size of this ellipse, our goal is to measure how bar formation affects the stellar kinematics. By keeping the measurement ellipse constant, our $\lambda_R$ measurements reflect the changing kinematics due to bar formation rather than due to a changing half mass ellipse. Also, by using a timestep where the bar has not formed, we avoid the issue of measuring the ellipticity of the bar rather than that of the galaxy. In both panels of Fig.~\ref{fig:massspin} and Fig.~\ref{fig:fluxspin}, the measurement ellipse of $i=60^\circ$ has a semi-major axis of 2.67 kpc with an ellipticity of 0.47, aligned with the horizontal axis. The measurement ellipse of $i=40^\circ$ and $i=80^\circ$ has semi-major axis 2.73 kpc and 2.49 kpc, respectively, with an ellipticity of 0.28 and 0.69, respectively.

In the top panel of Fig.~\ref{fig:massspin}, for each galaxy inclination, the evolution of the mass weighted spin appears to be similar. There is a period before bar formation when the galaxy spin is constant, a short period when there is a rapid decrease to a lower value and a period after where there is a small gradual decline. The latter period likely occurs due to the bar redistributing angular momentum during its secular evolution phase, causing $\lambda_R$ to slowly decrease over time. Also, higher inclinations of the galaxy result in $\lambda_R$ being larger, which is simply the azimuthal velocity component contributing more to the line of sight velocity. By taking the mean $\lambda_R$ before bar formation (t = 200 to 300 Myr) and after bar formation (t = 410 to 610 Myr), we find that the change in spin (fractional change) due to bar formation is roughly 0.04 (6.2\%), 0.04 (6.2\%) and 0.06 (8.6\%) for $i = 40^\circ$, $i = 60^\circ$ and $i = 80^\circ$ with uncertainties of 0.01, respectively.    

\begin{figure*}
	\includegraphics[width=\textwidth]{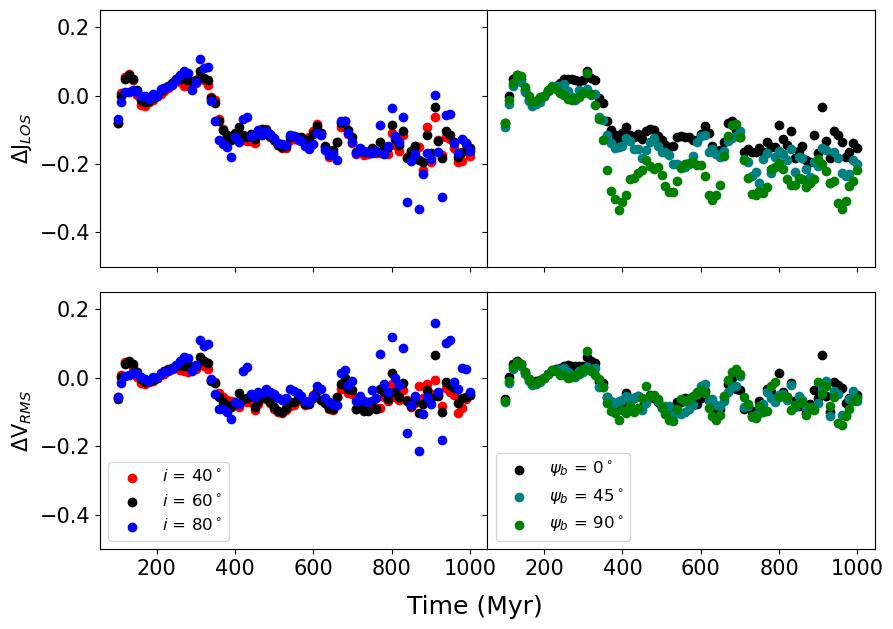}
    \caption{Fractional changes in the numerator ($J_{LOS}$) and denominator ($V_{RMS}$) of $\lambda_R$ normalized by the respective value at t = 200 Myr. The top left and bottom left panels show the change in $J_{LOS}$ and $V_{RMS}$, respectively, at inclinations $i = 40^\circ$, $i = 60^\circ$ and $i = 80^\circ$ with $\psi_b=0$. The top right and bottom right panels show the change in $J_{LOS}$ and $V_{RMS}$, respectively, at bar orientations $\psi_b = 0^\circ$, $\psi_b = 45^\circ$ and $\psi_b = 90^\circ$ with $i =60^\circ$.}
    \label{fig:numdnom}
\end{figure*}

It is somewhat surprising to see the small change in $\lambda_R$ before and after bar formation depend on the inclination of the galaxy. One tentative explanation could be that the change in $v_\phi$ we see in Fig.~\ref{fig:vel_disp_sim} becomes more apparent in the line of sight velocity at higher inclinations, causing a larger decrease. To examine this further, we plot the numerator ($J_{LOS}$) and the denominator ($V_{RMS}$) of the mass weighted $\lambda_R$:

\begin{equation}
    J_{LOS} = \sum_i^N M_iR_i|V_i|
    \label{eq:numer}
\end{equation}

\begin{equation}
    V_{RMS} = \sum_i^N M_iR_i\sqrt{V_i^2 + \sigma_i^2}
    \label{eq:denom}
\end{equation}

\noindent where $M$ is the stellar mass in the bin. The top left and bottom left panels of Fig.~\ref{fig:numdnom} show the fractional change of $J_{LOS}$ and $V_{RMS}$ at each inclination since their respective value at 200 Myr, $N_0$ and $D_0$. Changes in $J_{LOS}$ and $V_{RMS}$ are largely similar at each inclination, although there is more variance in both $J_{LOS}$ and $V_{RMS}$ at $i=80^\circ$. Therefore, inclination does not affect the measurement of $\lambda_R$ for a barred galaxy at any stage of its evolution. Any change we see can be attributed to noise. The different values of $\lambda_R$, seen the top panel of Fig.~\ref{fig:massspin}, can be corrected for by simply using edge-on galaxy spin, assuming this does not significantly increase the uncertainty. 

The bottom panel of Fig.~\ref{fig:massspin} shows the evolution of $\lambda_R$ of the bar angle ($\psi_b$) $0^\circ$, $45^\circ$ and $90^\circ$ from the horizontal axis. The inclination is kept fixed at $i=60^\circ$. Before bar formation, $\lambda_R$ is about the same for all three orientations as expected since there is no bar, but during bar formation the decrease in $\lambda_R$ is different for each of the orientations of the bar. Using the same time intervals as before to find the mean $\lambda_R$ (fractional change) before and after bar formation, we find the change to be 0.04 (6.2 \%), 0.07 (9.6 \%) and 0.11 (16 \%) for bar orientations of $0^\circ$, $45^\circ$ and $90^\circ$ with uncertainties of 0.01, respectively.

Analysis of the bottom panel of Fig.~\ref{fig:massspin} suggests that the more misaligned with the horizontal axis the major axis of the bar is, the bigger the change in $\lambda_r$ due to bar formation. We again plot the fractional change of $J_{LOS}$ (top right panel) and $V_{RMS}$ (bottom right panel) normalized by the value at 200 Myr for each $\psi_b$ in Fig.~\ref{fig:numdnom}. While the change in $V_{RMS}$ at $\psi_b = 90^\circ$ has larger variance over time, all three bar orientations fall by roughly similar amounts and follow the same trends. The reason for the trend in the change of $\lambda_r$ is seen in the top panel of Fig.~\ref{fig:numdnom}. For higher bar angles, bar formation leads to a greater decrease in $J_{LOS}$; about 12\%, 16\% and 23\% for $\psi_b$ $0^\circ$, $45^\circ$ and $90^\circ$, respectively. 

Eq.~\ref{eq:numer} is an analogue for the angular momentum within the measurement ellipse. If the overdensity of the bar is concentrated near the center of the ellipse, as it is at bar angle $90^\circ$, then the angular momentum within the ellipse will be low due to most of the mass tracing where the LOS velocity is low. The opposite case, where the bar is aligned with the ellipse will result in a higher angular momentum. This difference is what the top right panel of Fig.~\ref{fig:numdnom} shows, which is reflected in the change in $\lambda_R$.

In the top panel of Fig.~\ref{fig:vel_disp_sim}, we see the increase in velocity dispersion increase in all directions. However, the bottom panels of Fig.~\ref{fig:numdnom} show that $V_{RMS}$ decreases after bar formation. Therefore, the dominant driver of the change in $\lambda_R$ is the decrease in azimuthal velocity, seen in the bottom panel of Fig.~\ref{fig:vel_disp_sim} rather than the increase in velocity dispersion.

The expected change in mass weighted spin due to bar formation should be around 0.045 to 0.110, depending on $\psi_b$. While the change in galaxy spin may seem modest, it is significant as uncertainties in this parameter from IFU surveys like SAMI are typically around 0.01 (see Sec.~\ref{sec:samiexplain}).

\subsection{Effect of young stars on galaxy spin measurements}
\label{sec:fluxspinbarform}

In Sec.~\ref{sec:galaxyspinbarform} we explored changes in mass weighted $\lambda_R$, however $\lambda_R$ is typically measured using a flux weighting. Since Fig.~\ref{fig:agesfr_sim} shows bar formation leads to a burst of star formation in the bar region, the kinematics of these younger, brighter stars may dominate the flux weighted $\lambda_R$. 

\begin{figure}
	\includegraphics[width=\columnwidth]{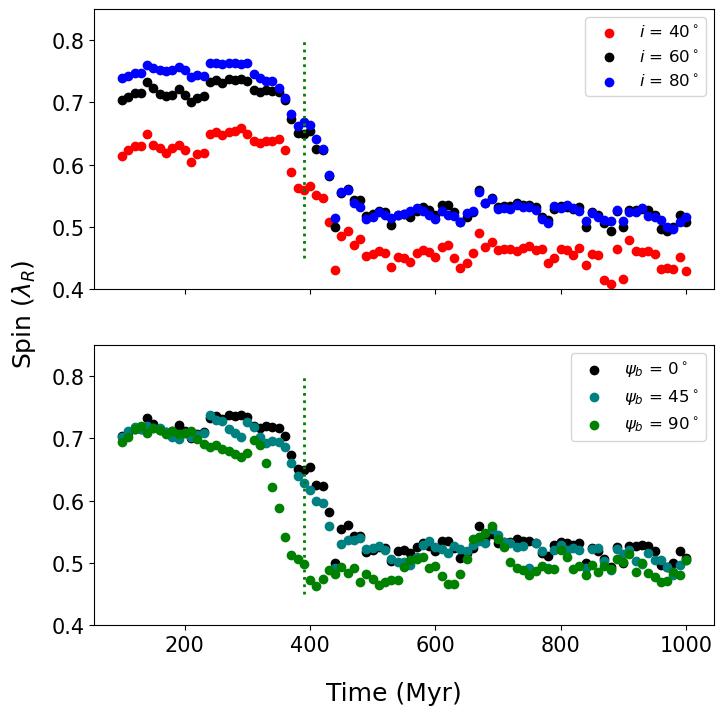}
    \caption{The evolution of r-band flux weighted galaxy spin within a 1 $R_e$ ellipse measured at t = 200 Myr. Top panel - at inclinations of $i = 40^\circ$, $i = 60^\circ$ and $i = 80^\circ$ with the bar aligned with the horizontal axis. Bottom panel - at a fixed inclination of $i = 60^\circ$ with the orientation of the bar with respect to the horizontal axis at $\psi_b = 0^\circ$, $\psi_b = 45^\circ$ and $\psi_b = 90^\circ$. The dotted green line indicates when the bar forms. }
    \label{fig:fluxspin}
\end{figure}

Fig.~\ref{fig:fluxspin} is similar to Fig.~\ref{fig:massspin}, except now we use the r-band flux from \textsc{simspin} (see Sec.~\ref{sec:simspin}) to measure galaxy spin. In the top panel of Fig.~\ref{fig:fluxspin}, the flux weighted $\lambda_R$ initially follows a similar trend to the mass weighted $\lambda_R$, in the sense that higher inclinations result in higher values of spin, although the difference in $\lambda_R$ at the given inclinations is now lesser than it was in the top panel of Fig.~\ref{fig:massspin}. Due to the smaller initial difference in $\lambda_R$ in the top panel of Fig.~\ref{fig:fluxspin}, $\lambda_R$ at $i = 60^\circ$ and $i = 80^\circ$ converge after bar formation. However, this is similar to the trend found in the top panel of Fig.~\ref{fig:massspin}, where at $i = 60^\circ$ and $i = 80^\circ$ the difference in $\lambda_R$ at these inclinations decreased during and after bar formation. The surprising difference is that the decrease in flux weighted $\lambda_R$ during bar formation continues for about 100 Myr after the bar forms, in contrast to the top panel of Fig.~\ref{fig:massspin} where the decrease stops right when the bar forms.

The trend found in the bottom panel of Fig.~\ref{fig:fluxspin} is different altogether for each bar angle than the trend found in Fig.~\ref{fig:massspin}. The decrease in $\lambda_R$ continues after bar formation, although for the bar angle of $90^\circ$ the decrease stops shortly after bar formation. Also worth noting is the fact that, for all three bar orientations, the mean $\lambda_R$ after the decrease is around the same value, somewhere between 0.50 and 0.55.

\begin{figure}
	\includegraphics[width=\columnwidth]{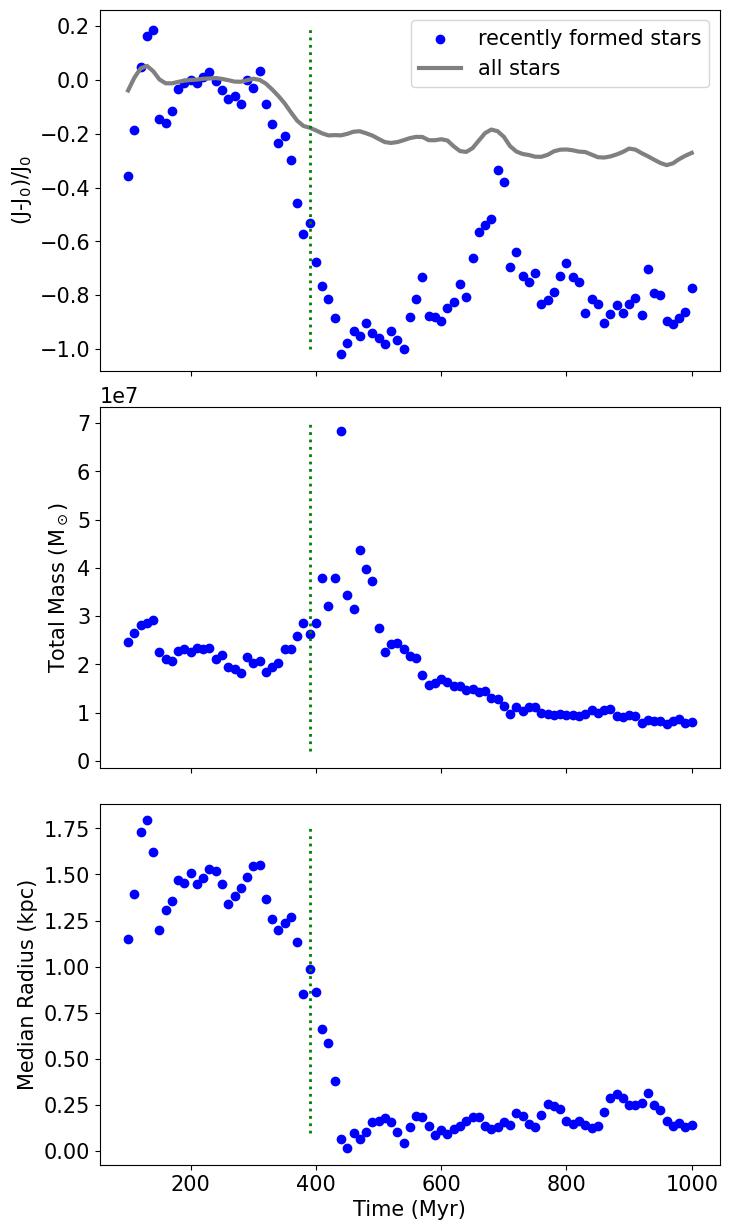}
    \caption{Properties of stars formed in the last 10 Myr at each simulation snapshot within the bar region (blue points). Top panel - fractional change in specific angular momentum ($J$) with respect to the value at t = 200 Myr. The grey line shows the same quantity using all stars in bar region. Middle panel - The total mass of the newly formed stars in bar region. Bottom panel - the mean radius of the newly formed stars in bar region. The dotted green line indicates bar formation time.}
    \label{fig:newstars}
\end{figure}

To understand the differences between Fig.~\ref{fig:massspin} and Fig.~\ref{fig:fluxspin} we need to examine the properties of newly formed stars at each snapshot within the bar region. Newly formed stars are defined as stars born in the last 10 Myr from the time of the snapshot. The top panel of Fig.~\ref{fig:newstars} shows the fractional change of specific angular momentum in the z direction ($J$) since the 200 Myr snapshot ($J(200) = J_0$). Using all stars within the bar region (grey line in top panel of Fig.~\ref{fig:newstars}) to calculate the change in $J$ we see a sharp decline of about 20\% during the bar formation period. After the bar forms there is a gradual, but admittedly small, decrease associated with the secular evolution of a bar. Tracking $J$ of the newly formed stars (blue points in top panel of Fig.~\ref{fig:newstars}) shows that due to bar formation, these stars have significantly lower angular momentum than stars formed before bar formation. This trend seems to continue until the end of the simulation.

The star formation rate shown in Fig.~\ref{fig:agesfr_sim} indicates the degree to which the kinematics of these young stars will dominate the flux weighted $\lambda_R$ measurement. To further illustrate this we plot the total mass of all new stars within the bar radius at each time in the middle panel of Fig.~\ref{fig:newstars}. The bottom panel of Fig.~\ref{fig:newstars} shows the mean radius of new stars born in the bar region. These two panels show that right after the bar forms there is an increase in the number of new-born stars which increases the total stellar mass at the center of the galaxy. This can be seen visually using snapshots of the simulation after the starburst in the bright central pixels in panels D and E of Fig.~\ref{fig:faceon_maps_sim}. In our simulation, the new-born stars at the center of the galaxy accounts for more than roughly 60\% of the total flux in the measurement ellipse as can be seen in the top-right panel of Fig.~\ref{fig:weakstrongexamples}. These low $J$ stars are born near or at the center of the galaxy, forming the commonly seen nuclear disc seen in simulations of barred galaxies \citep{nucleardisc2014}. Therefore, the decrease in $\lambda_R$ we see in Fig.~\ref{fig:fluxspin} is a result of these new stars dominating in flux, which means they contribute more than the older stars to $\lambda_R$. Since most of these stars are at or near the center changing the bar angle no longer has a major effect on $\lambda_R$ measurements.

It is important to note that results from numerical studies of star formation induced by bar formation are different depending on the simulations used. For example, \citet{barquenching2018khop} showed that star formation decreases after bar formation since the gas is too dynamically hot to collapse and form stars. In contrast, \citet{barstarburstcarles2016} find that bar formation leads to a starburst event. In our work we see a large central starburst, also examined in the work of \citet{lizsimuation2022}, which dominates the flux-weighted $\lambda_{Re}$ measurement. However, this is highly dependent on the recipe of star formation used in the simulation. Therefore, the change in $\lambda_R$, seen in both panels of Fig.~\ref{fig:fluxspin} of about 0.2 is likely extreme and occurs due to the dramatic difference in flux from newly formed stars and the initial condition stars. Therefore, from this study, we can anticipate star formation due to bar formation to increase the expected change range found in Sec.~\ref{sec:galaxyspinbarform} of around 0.04 to 0.11 in a qualitative sense. A comprehensive look on how star formation plays a role in $\lambda_R$ measurements is left for future studies.    

\subsection{Connection with SAMI data}
\label{sec:samidataresults}

\begin{figure}
	\includegraphics[width=\columnwidth]{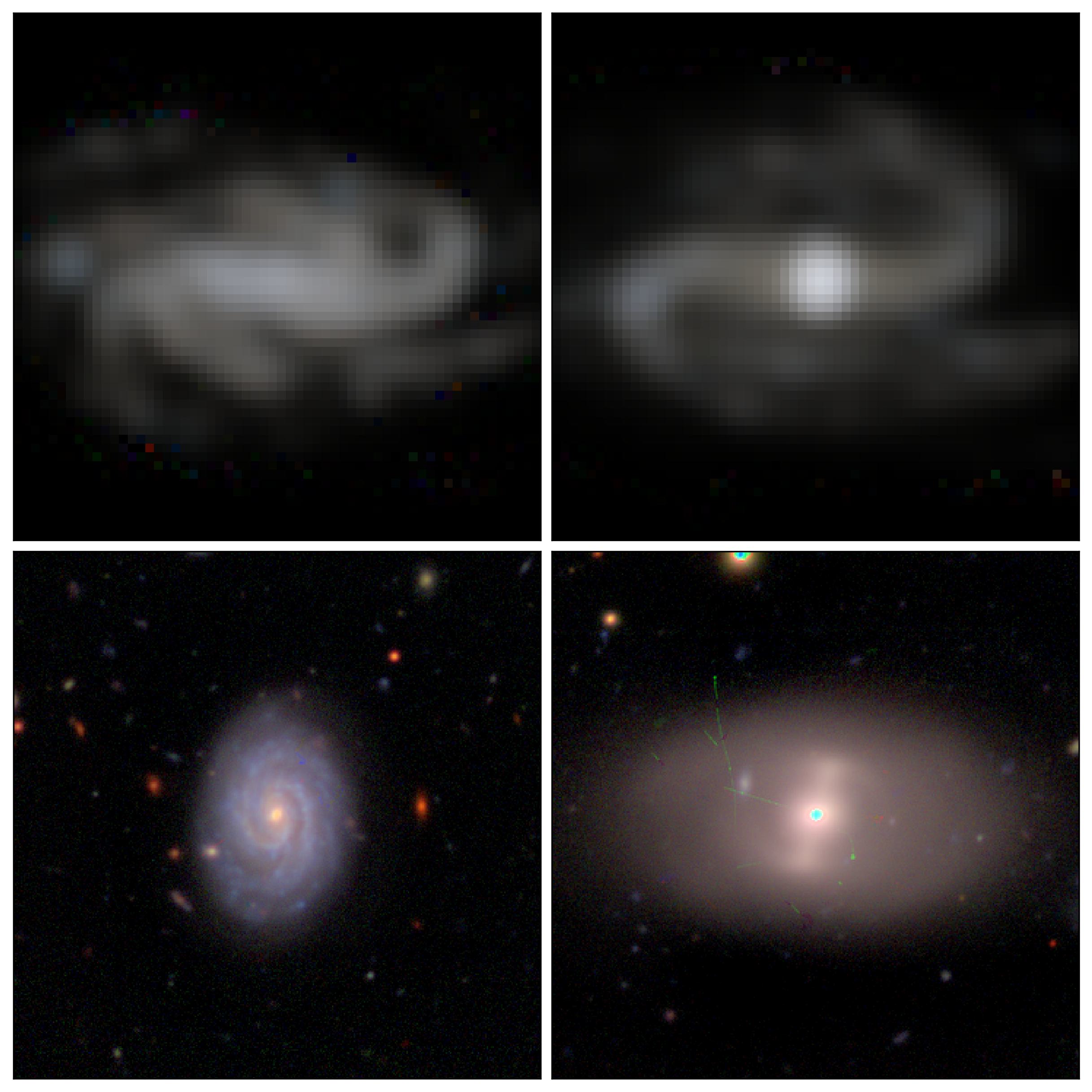}
    \caption{Examples of weak and strong bars from our simulation and HSC-SSP imaging. Top left - a color image of the 330 Myr snapshot using g,i and r band flux from \textsc{simspin} inclined at $i=60^\circ$. Top right - a color image of the 600 Myr snapshot using g,i and r band flux from \textsc{simspin} inclined at $i=60^\circ$. Bottom left - HSC-SSP image of a weak bar in SAMI. Bottom right - HSC-SSP image of a strong bar in SAMI.}
    \label{fig:weakstrongexamples}
\end{figure}

\begin{figure}
	\includegraphics[width=\columnwidth]{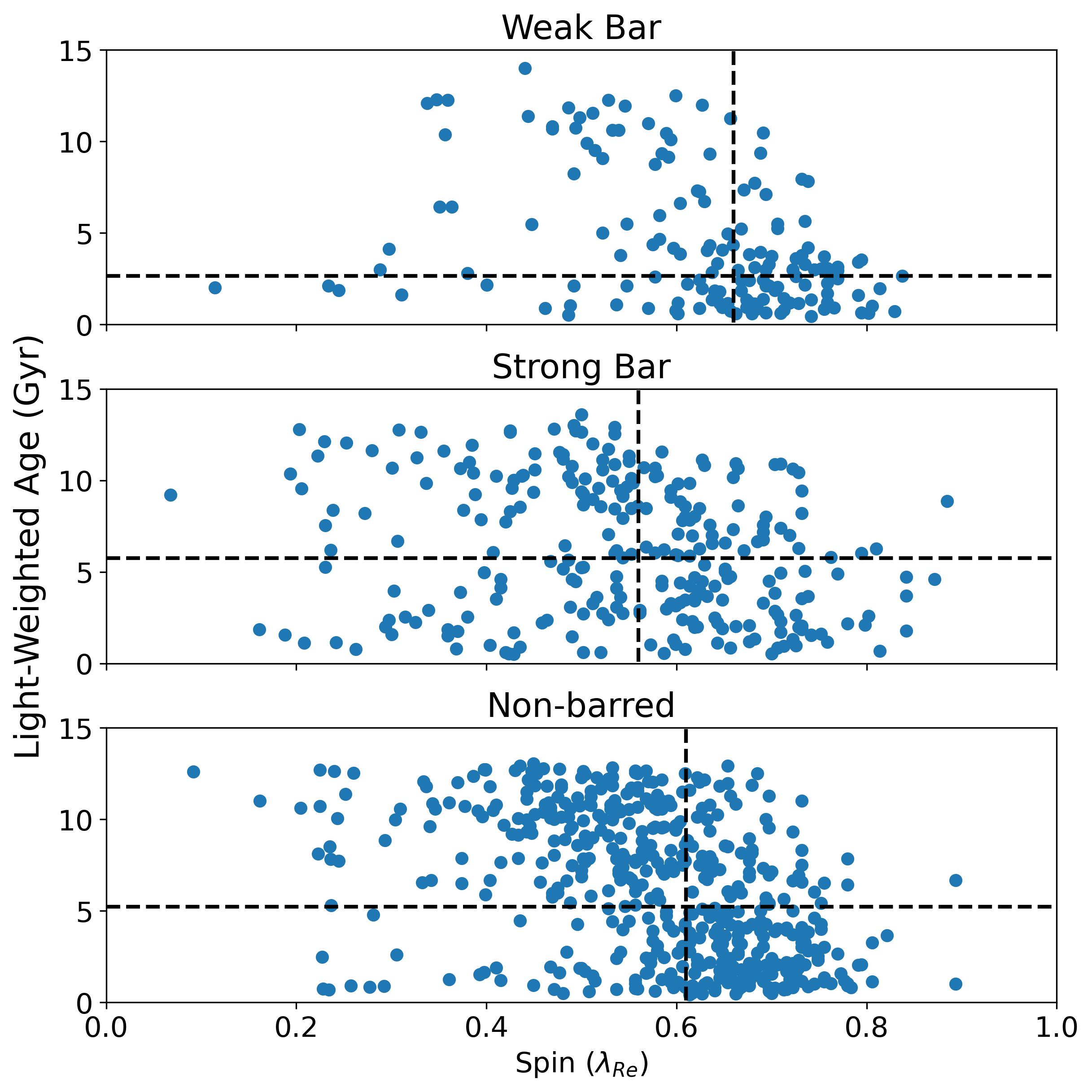}
    \caption{Two-dimension distribution of galaxies in light weighted stellar population age versus spin space. The median value of spin and median age are indicated by the vertical and horizontal, dashed black line. Top panel: weakly barred galaxies. Middle panel: Strongly barred galaxies. Bottom panel: Non-barred galaxies.}
    \label{fig:age_spin_sami}
\end{figure}

Our goal is to connect what we know about bar formation in simulations to the measured stellar kinematics provided by surveys like SAMI. To do this, we compare the properties of strongly barred, weakly barred and non-barred galaxy populations described in Sec.~\ref{sec:samiexplain}. In Fig.~\ref{fig:weakstrongexamples}, the bottom left panel is a representative weakly barred galaxy and the bottom right panel a strongly barred galaxy. These images were made using the g, r and i band maps taken from the Hyper-Suprime-Cam Subaru Strategic Program's (HSC-SSP \citep[HSC2018][]{HSC2018}) public data release 2 \citep{hscdr2}. The top left and right panels are examples of weak and strong bars, respectively, from our simulation using g,r, and i band flux maps from \textsc{simspin}.    

The distribution of weakly barred galaxies in light-weighted age and spin within 1 $R_e$ is shown in the top panel of Fig.~\ref{fig:age_spin_sami}. Immediately obvious is the high density of galaxies that have low light-weighted age and high galaxy spin. About 58\% of the weakly barred galaxy sample have age less than 5 Gyr with spin greater than 0.6. By contrast, shown in the middle panel and bottom panel of Fig.~\ref{fig:age_spin_sami} is the distribution of strongly barred and non-barred galaxies. Galaxies in both of these groups are visually more evenly distributed with a trend of older galaxies having lower spin. 

To further illustrate the difference between each population of galaxy we show the median values of light-weighted age and $\lambda_{Re}$ with the dotted black lines in Fig.~\ref{fig:age_spin_sami}. The median spin and inter-quartile range (IQR) of the weakly barred sample is 0.66 (0.58 - 0.73), 0.56 (0.48 - 0.65) for the strongly barred sample and 0.61 (0.51 - 0.69) for the non-barred sample. While there is some overlap between the different distributions, the Anderson-Darling test \citep{AndersonDarling} indicates that the weakly barred sample distribution is significantly different than the strongly barred and non-barred sample at the $0.5\%$ significance level in both $\lambda_{Re}$ and light-weighted age. The strongly barred and non-barred sample are also significantly different.

\begin{figure}
	\includegraphics[width=\columnwidth]{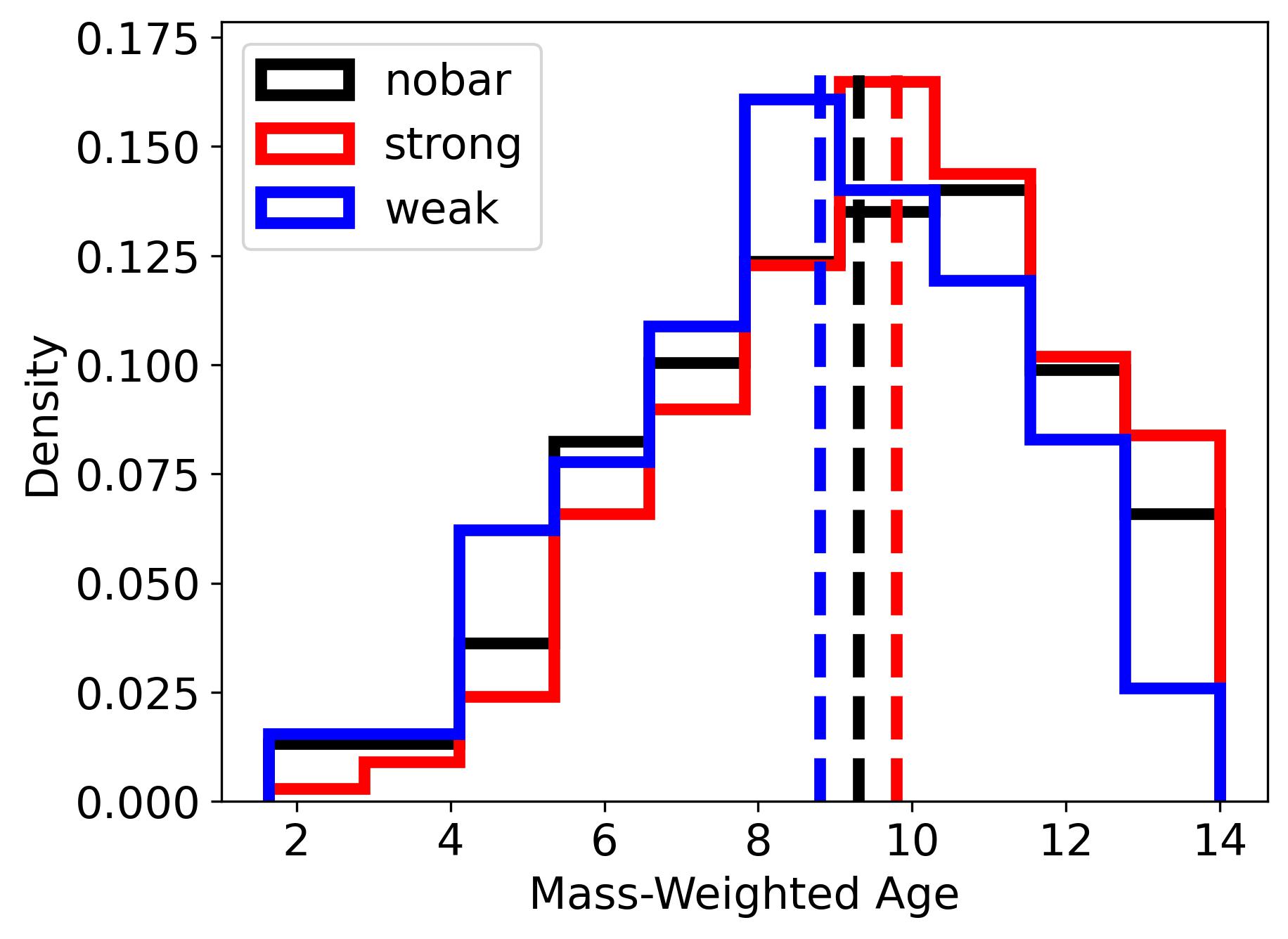}
    \caption{The distribution of mass-weighted ages for weakly barred, strongly barred and non-barred galaxies within one $R_e$. The dashed lines indicate median values}
    \label{fig:samimassage}
\end{figure}

The distribution of mass-weighted ages among non-barred, weakly and strongly barred galaxies also show some differences. In Fig.~\ref{fig:samimassage}, the distribution of each sample is shown with the median values indicated by dashed lines. The median mass-weighted age (IQR) for non-barred galaxies is 9.3 (7.3-11.1), for strongly barred galaxies is 9.8 (7.9 - 11.3) and 8.8 (7.0 - 10.6) for weakly barred galaxies. Only the strongly and weakly barred samples are significantly different using the Anderson-Darling test at the 0.5\% significance level.

\begin{figure}
	\includegraphics[width=\columnwidth]{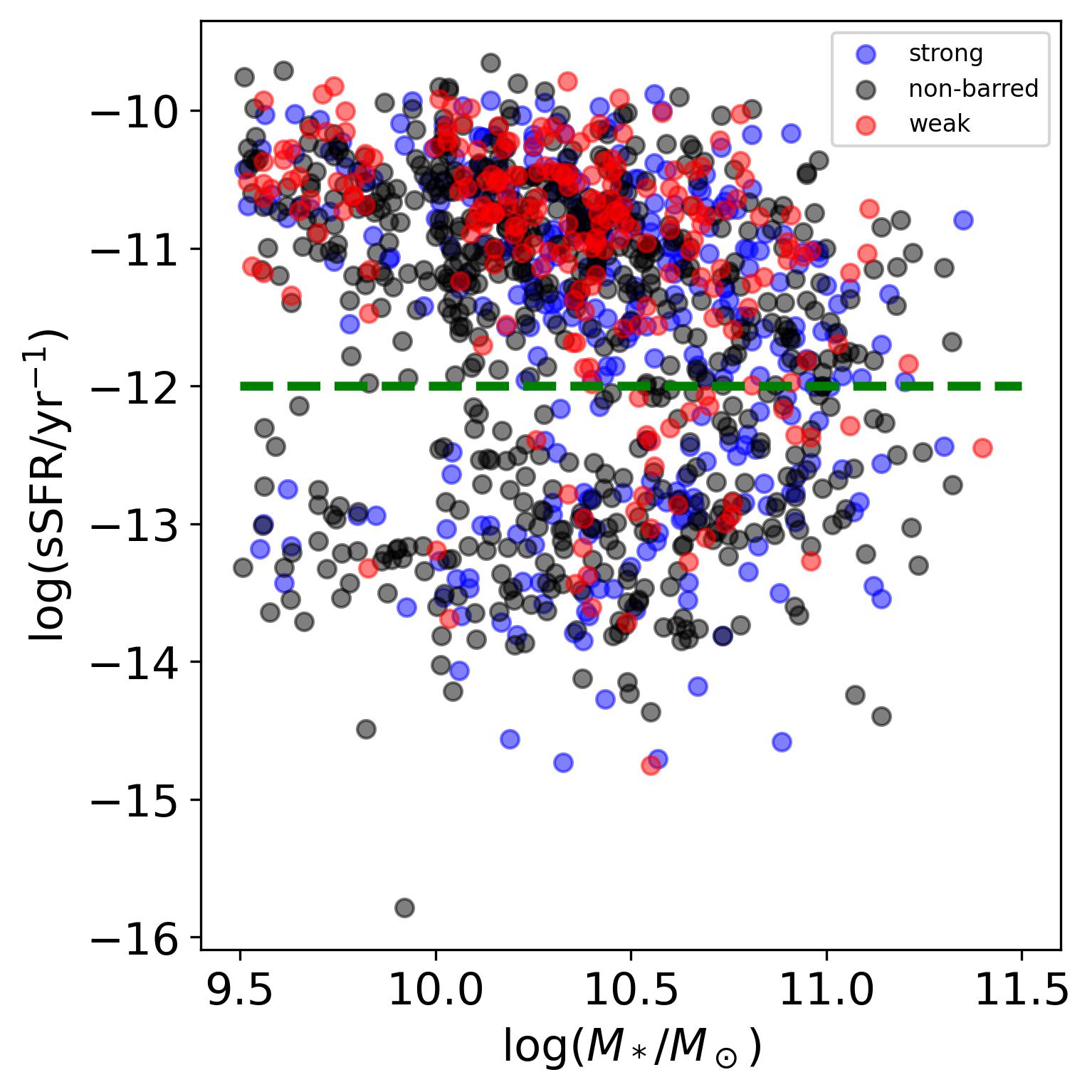}
    \caption{The distribution of weakly barred, strongly barred and non-barred galaxies by log stellar mass and log specific star formation rate (sSFR) within one $R_e$. The dashed green line indicates where we define galaxies to be star forming or quenched.}
    \label{fig:samimassssfr}
\end{figure}

In Sec.~\ref{sec:heatingsfr}, we found that after bar formation there is a short starburst period after which the SFR drops to less than half of the SFR when the bar was still growing. Fig.~\ref{fig:samimassssfr} shows the distribution of galaxies of each population by log stellar mass and log specific star formation rate (sSFR) within $1R_e$. Noteworthy is the different distribution in sSFR for each sample. If we consider galaxies above $\log(\text{sSFR}/\text{yr}^{-1}) = -12$ to be star forming, then 84\% of weakly barred, 68\% of strongly barred and 64\% of non-barred galaxies lie above this line. The median (IQR) $\log(\text{sSFR}/\text{yr}^{-1})$ for weakly barred galaxies are -10.8 (-11.3 - (-10.5)), -11.3 (-12.6 - (-10.7)) for strongly barred and -11.3 (-12.7 - (-10.7)) for non-barred galaxies. While there is overlap between the three distributions, the weakly barred distribution is significantly different in sSFR from both strong and non-barred distributions using the Anderson-Darling test, at the $0.5\%$ significance level.

There are also some differences between the distributions of stellar mass shown in Fig.~\ref{fig:samimassssfr}. The non-barred galaxies tend to have lower $\log(M_*/M_\odot)$ with the median (IQR) being 10.32 (10.04 - 10.63). Strongly barred galaxies tend to have the most massive discs with median (IQR) 10.42 (10.20 - 10.69) and weakly barred galaxies sit in between with median (IQR) 10.37 (10.12 - 10.60). Applying the Anderson-Darling test at the $0.5\%$ significance level shows only the strongly barred and non-barred distributions to be significantly different. A possible interpretation of this trend can be that the baryons of heavier discs are more dominant over the dark matter halo at the center of the galaxy compared to lighter discs, which would cause a faster bar formation timescale \citep{joss2023fdisk}. However, more data is needed to definitively confirm this result, something we leave for future works.

\section{Discussion}
\label{sec:discussion}

\subsection{What are weak and strong bars?}
\label{sec:strongweakexplain}

Analysis of our simulation indicates that the formation of a bar in a cold rotating disc will cause a decrease in $\lambda_R$, the magnitude of which depends on the orientation of the bar in the galaxy and the star formation due to bar induced gas inflow. Since the drop in $\lambda_R$ begins around the time when the bar begins to form, one would expect the strength of the bar to be anti-correlated with $\lambda_R$. However, it is important to note that our simulation only explores the formation and early secular evolution stages of the bar. After bar formation, the bar enters its secular evolution phase and redistributes angular momentum to the outer disk and dark matter halo \citep{athabarsecular} which increases bar strength (briefly mentioned in Sec.~\ref{sec:galaxyspinbarform}). The rate of angular momentum loss in the inner disc depends on the properties of the dark matter halo such as halo spin \citep{bardarkcollier2021,barangkararia2022,baranglu2024joshi} or the presence of a spherical bulge \citep{bulgesbarsimkat2018}. 

The results using galaxies from the SAMI galaxy survey show that weakly barred galaxies have lower mass-weighted and light-weighted ages, higher sSFR and higher $\lambda_{Re}$ values than strongly barred galaxies. One can interpret the higher sSFR and lower light-weighted age in weakly barred galaxies as indications of a more active galaxy, while, as our simulation shows, strongly barred galaxies use up their star-forming gas rapidly due to the strong torque that is applied by the strong bar, leading to inert galaxies. An interpretation of the lower mass-weighted age is that weakly barred galaxies have younger stellar discs than strongly barred galaxies, suggesting that weakly barred galaxies are simply less evolved than strongly barred galaxies. Should a weakly barred galaxy evolve in isolation into a strongly barred galaxy, then $\lambda_{Re}$ and the sSFR would drop, as we find in both our simulation and the SAMI sample. 

It is worth noting that the distribution of non-barred galaxies consists primarily of two types of galaxies: older, low $\lambda_{Re}$ S0s and young, high $\lambda_{Re}$ late-types. The results we have reported until now have included the S0s in this sample, however the formation of S0s occurs through the stripping of gas or mergers \citep{s0formationdeeley2021}, which are processes that are external to a galaxy. Once these galaxies are removed the median light-weighted age decreases from around 5 Gyr to 4 Gyr and $\lambda_{Re}$ increases from 0.61 to 0.64. The median mass-weighted age also decreases to 8.8 Gyr. The Anderson-Darling test at the 0.5\% significance level shows the non-barred and weakly barred distributions to no longer be significantly different in light-weighted age and $\lambda_{Re}$. This is in line with our findings since the remaining galaxies in the non-barred sample, after removing S0s, are rotating cold discs.   

In Sec.~\ref{sec:samidataresults} we find that the median $\lambda_{Re}$ for weakly barred galaxies is higher than for strongly barred galaxies, which supports what our simulation shows. However, the observed median light-weighted age for strongly barred galaxies is around $ 6$ Gyr while for weakly barred galaxies it is around $3$ Gyr. This indicates that the strength of the bar depends primarily on light-weighted age, therefore the difference in $\lambda_{Re}$ measurement may be driven by the secular evolution of the bar rather than bar formation. Of course, other age-specific processes, such as early turbulent star formation may also play a role. 

\begin{figure}
	\includegraphics[width=\columnwidth]{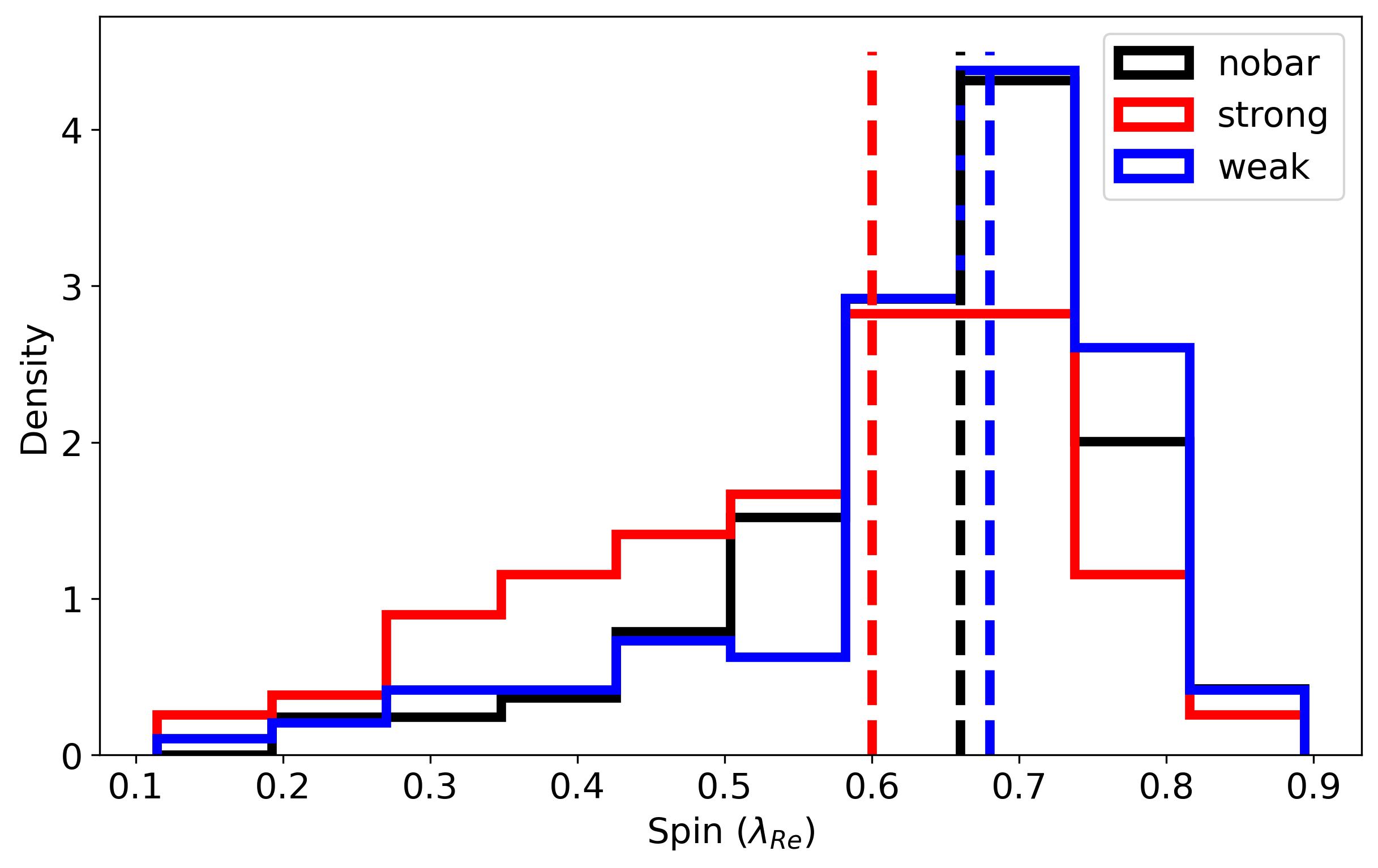}
    \caption{Distribution of strongly, weakly and non-barred galaxies by $\lambda_{Re}$ for SAMI galaxies with age less than 3 Gyr. The dashed lines indicate median values. }
    \label{fig:samiyoungspin}
\end{figure}

To further test whether the affect of bar formation on $\lambda_R$ seen from our simulation can be seen in SAMI data, we need to isolate galaxies where there is no expected change in $\lambda_{Re}$ due to affects related to light-weighted age. \citet{croomAgespin2024} found that there is no obvious decline in $\lambda_{Re}$ with a full SAMI galaxy sample between light-weighted ages up to 3 Gyr. Therefore, by examining galaxies in this age range we can isolate for the effect of bar formation on $\lambda_{Re}$. 

In Fig.~\ref{fig:samiyoungspin} we look at the distribution of the three populations of galaxies from SAMI with light-weighted ages less than 3 Gyr. There are 123 weakly barred, 100 strongly barred and 211 non-barred galaxies in this sample. Of the 100 strongly barred galaxies only 18 have BP bulges (although this classification depends on the inclination which is discussed in Sec.~\ref{sec:bias}). The distribution of $\lambda_{Re}$ in the non-barred and weakly barred galaxies is not significantly different from each other but is significantly different from the strongly barred galaxies using an Anderson-Darling test at the 0.5\% significance level. The median $\lambda_{Re}$ for strongly barred galaxies is about 0.08 and 0.06 less than weakly and non-barred galaxies, respectively, falling within the range we find in Sec.~\ref{sec:galaxyspinbarform}. This can be interpreted as non-barred and weakly barred galaxies having colder discs while strongly barred galaxies have been dynamically heated by the bar formation process, as we find in our simulation. Furthermore, we find the mass distribution of this weak and strong bar sample to not be significantly different from each other using the Anderson-Darling test with median values around $\log(M_*/M_\odot) = 10.2$. These findings show $\lambda_{Re}$ to be a useful parameter in determining the kinematic and morphological properties of barred galaxies.

It is worth considering our findings in the broader context of how older galaxies seem to have lower values of $\lambda_{Re}$ \citep{croomAgespin2024}. To thoroughly study whether the bar alone can spin down a galaxy to the extent seen in \citet{croomAgespin2024}, one would need to explore the evolution of a barred galaxy over a longer period of time than what we explore in this study, hence we leave this for future works. However, the \citet{croomAgespin2024} trend can be seen visually in Fig.~\ref{fig:age_spin_sami}, where for all three distributions, older galaxies have a lower galaxy spin. Also, considering the median light-weighted age and $\lambda_{Re}$ for strong and weak bar galaxies, one interpretation that is supported by our simulation is that the bar plays a role in driving $\lambda_{Re}$ down over time.  

Along with kinematic properties, the $\log(\text{sSFR}/\text{yr}^{-1})$ also seems to be different for the three galaxy populations, as seen in Sec.~\ref{sec:samidataresults}. Our findings agree with those of \citet{weakstrongbarstarformvera2016}, who found that strongly barred galaxies have lower star formation activity and older stellar populations than weakly barred galaxies. However, if we only consider galaxies younger than 3 Gyr, the distribution of all three populations are not significantly different with median values -10.51, -10.49 and -10.57 for weak, strong and non-barred galaxies, respectively. 

It is important to note that $\log(\text{sSFR}/\text{yr}^{-1})$ values within 1 $R_e$ do not contain any information on the location of star formation. \citet{starformlocgeron2024} found that strongly barred galaxies from the MaNGA galaxy survey have enhanced star formation in the center compared to weakly and non-barred galaxies, suggesting weak bars do not cause significant gas inflows. Alternatively, gas inflows may be delayed due to the weak torque applied by weak bars. This result is in agreement with our simulation, but the mapping of star formation spatially in SAMI galaxies is beyond the scope of this work.

We have so far only considered bar formation driven by internal mechanisms, but tidally formed bars have not been considered. Bars formed through tidal interactions tend to be stronger \citep{tidalbarsstronger2019} and rotate more slowly \citep{tidalslowrotate2017} than bars formed in isolation. Recently, \citet{barformationtidalshen2025} used N-body simulations to find that if a disc is inherently bar unstable, bar formation will be driven by swing amplification, regardless of whether there is a tidal interaction. However, they also find that if a disc is stable to the bar instability, then a tidal interaction can drive bar formation through an alternative mechanism. This idea was recently explored using the TNG50 cosmological simulation by \citet{tidalbarscosmo2025}, who found that about 25\% of barred galaxies in their sample of 307 formed a bar due to tidal interactions that are otherwise bar stable. Recently, \citet{spinMANGAsodi2026} suggested the internal bar formation mechanism to be the primary cause of bars in the MaNGA sample of barred galaxies, hence it is appropriate to consider the weak and strong bar dichotomy in this paradigm as we have done. Further work is needed to understand these topics in more detail, such as the stellar kinematics of galaxies with a bar formed through tidal interactions of otherwise stable discs.

\subsection{Classifying Bars and Determining \boldmath{$\lambda_{Re}$}}
\label{sec:bias}

While we have shown $\lambda_{Re}$ to change depending on the classification of barred galaxies, the accuracy of this parameter depends greatly on the measurement ellipse. The ellipticity of SAMI galaxies is determined by the best fitting MGE model within 1 $R_e$ \citep{jessespin2017,d'eugenioremeasurement2021}. In the presence of a relatively strong bar, the measured ellipticity will be that of the bar rather than that of the galaxy itself. In our idealized simulation study we avoid this issue entirely by using a fixed ellipse drawn before the formation of the bar. By contrast, with observations from surveys like SAMI, especially for strongly barred galaxies, it is possible for the ellipticity to be that of the bar rather than that of the galaxy. This would make the edge-on spin, which is needed to correct for the effect of inclination, not reflect the true projection of these galaxies. Future works using the galaxy spin for bars should take steps, such as visual inspection, to ensure this is not the case or by using outer ellipses that better define the disc shape.          

There may also be biases in the visual classifications of bars. First, it is difficult to classify a galaxy as barred if the inclination is close to edge-on without the presence of a BP bulge. In this study we classify BP bulges as strong bars, as discussed in Sec.~\ref{sec:samiexplain}, so this would mainly affect young strong bars and weak bars. This subset of galaxies, presumably small, would be classified as non-barred galaxies in this study. It is also important to note that the visual classification of weak and strong is not quantitative and depends greatly on the judgment of those classifying these images. The use of $\lambda_{Re}$ in this regard is to quantify how much angular momentum the bar has redistributed from the center of the stellar disc, which as our results show, tracks well with the weak and strong dichotomy. Other classification methods, such as deep learning models, have also been used by \citet{machinelearningclassification2024} to measure the bar length of SAMI galaxies. They find bars with low $\lambda_{Re}$ tend to have longer bar lengths (normalized by host galaxy scale length), which is in agreement with our findings.

\section{Conclusion}
\label{sec:conclusion}

In this study we examined how bar formation affects $\lambda_R$ and how this parameter changes depending on the weak or strong classification of barred galaxies. We conducted analysis on a simulation of an isolated galaxy that forms a bar to confirm keys results related to how bar formation occurs and changes star formation, velocity dispersion and azimuthal velocity. Then, making use of \textsc{simspin} we examine how $\lambda_R$ is affected by bar formation and the orientation of the galaxy. Finally, we examine barred galaxies in the SAMI survey and drawn parallels to results from our simulation. 

The key takeaways from this work are as follows:

\begin{itemize}
    \item Bar formation will decrease $\lambda_R$ due to the redistribution of angular momentum resulting in the change of kinematics caused by this process. At a constant inclination, the decrease in mass-weighted $\lambda_R$ ranges from 0.04 to 0.11 depending on the the orientation of the bar in the galactic disc. If there is a starburst due to bar driven gas inflows then this decrease may be larger due to new stars at the center of the galaxy dominating the flux map. 

    \item By comparing the distribution of strongly, weakly and non-barred SAMI galaxies we find that weak bars tend to have higher $\lambda_{Re}$ values and are generally younger in both mass-weighted and light-weighted age than strongly barred galaxies. The $\log{\text{sSFR}/\text{yr}^{-1}}$ distribution of weakly barred galaxies is significantly different than those of strongly barred and non-barred galaxies, with weakly barred galaxies having higher star formation than strongly and non-barred galaxies.

    \item Since the median light-weighted age of weakly barred galaxies is about 3 Gyr lower than strongly barred galaxies, a large driver of the difference in $\lambda_{Re}$ might be the redistribution of angular momentum by the bar during secular evolution rather than bar formation itself. But further isolating those galaxies that have light-weighted stellar population age less than 3 Gyr, we find that strongly barred galaxies still have a lower galaxy spin than weakly and non-barred galaxies. This may suggest that many weak bars are in the bar formation phase (growing rapidly through swing amplification), while strong bars are in the secular evolution phase.
\end{itemize}

There are many ways to build on this work in future studies. New insights on bar formation and evolution are emerging from high-resolution simulations that consider a wide range of gas processes \citep{joss2024turb,joss2025slosh}.
Another is to re-examine the trend found in this study with data from the next generation of integral field spectroscopy. For example, the Hector survey \citep[][successor to SAMI]{Hector2024} aims to gather data on about 15000 galaxies with higher spatial coverage and spectral resolution which may be valuable in directly seeing the impact of the bar in stellar kinematics. Along with this, if visual classifications of bars can be made quantitative, using new high-resolution imaging data from Euclid \citep{euclid2025}, then analysis using stellar kinematics can become a key method in testing predictions made by idealized and cosmological simulations.     

\section*{Acknowledgements}

We thank Kate Harborne for help with \textsc{simspin} and Tomas H. Rutherford for help with the HSC-SSP images. RJ acknowledges support from the University of Sydney's Physics Foundation Scholarship. EJI acknowledges the support of the Australian Research Council through Discovery Project DP220103384.

The SAMI Galaxy Survey is based on observations made at the Anglo-Australian Telescope. The Sydney-AAO Multi-object Integral field spectrograph (SAMI) was developed jointly by the University of Sydney and the Australian Astronomical Observatory. The SAMI input catalogue is based on data taken from the Sloan Digital Sky Survey, the GAMA Survey, and the VST ATLAS Survey. The SAMI Galaxy Survey is supported by the Australian Research Council Centre of Excellence for All Sky Astrophysics in 3 Dimensions (ASTRO 3D), through project number CE170100013, the Australian Research Council Centre of Excellence for All-sky Astrophysics (CAASTRO), through project number CE110001020, and other participating institutions. The SAMI Galaxy Survey website is http://sami-survey.org/.

The Hyper Suprime-Cam (HSC) collaboration includes the astronomical communities of Japan and Taiwan, and Princeton University. The HSC instrumentation and software were developed by the National Astronomical Observatory of Japan (NAOJ), the Kavli Institute for the Physics and Mathematics of the Universe (Kavli IPMU), the University of Tokyo, the High Energy Accelerator Research Organization (KEK), the Academia Sinica Institute for Astronomy and Astrophysics in Taiwan (ASIAA), and Princeton University. Funding was contributed by the FIRST program from Japanese Cabinet Office, the Ministry of Education, Culture, Sports, Science and Technology (MEXT), the Japan Society for the Promotion of Science (JSPS), Japan Science and Technology Agency (JST), the Toray Science Foundation, NAOJ, Kavli IPMU, KEK, ASIAA, and Princeton University. 

This paper makes use of software developed for the Large Synoptic Survey Telescope. We thank the LSST Project for making their code available as free software at  http://dm.lsst.org

The Pan-STARRS1 Surveys (PS1) have been made possible through contributions of the Institute for Astronomy, the University of Hawaii, the Pan-STARRS Project Office, the Max-Planck Society and its participating institutes, the Max Planck Institute for Astronomy, Heidelberg and the Max Planck Institute for Extraterrestrial Physics, Garching, The Johns Hopkins University, Durham University, the University of Edinburgh, Queen’s University Belfast, the Harvard-Smithsonian Center for Astrophysics, the Las Cumbres Observatory Global Telescope Network Incorporated, the National Central University of Taiwan, the Space Telescope Science Institute, the National Aeronautics and Space Administration under Grant No. NNX08AR22G issued through the Planetary Science Division of the NASA Science Mission Directorate, the National Science Foundation under Grant No. AST-1238877, the University of Maryland, and Eotvos Lorand University (ELTE) and the Los Alamos National Laboratory.

Based [in part] on data collected at the Subaru Telescope and retrieved from the HSC data archive system, which is operated by Subaru Telescope and Astronomy Data Center at National Astronomical Observatory of Japan.

\section*{Data Availability}

Data will be made available upon reasonable request. Analysis was conducted using original \textsc{python} code using the \textsc{NumPy} \citep{numpy2020}, \textsc{SciPy} \citep{scipy2020}, \textsc{AstroPy} \citep{astropy:2013,astropy:2018,astropy:2022}, \textsc{pandas} \citep{pandas2010}, \textsc{Matplotlib} \citep{matplotlib2007}, \textsc{pynbody} \citep{pynbody} and \textsc{pytreegrav} \citep{pytreegrav} libraries. 


\bibliographystyle{mnras}
\bibliography{example} 

\bsp	
\label{lastpage}
\end{document}